\documentclass[preprint2]{aastex6}
\renewcommand{\baselinestretch}{1.0}
\usepackage{graphicx}
\usepackage{placeins}
\usepackage{xcolor}
\usepackage{natbib}
\usepackage{aasref}
\begin{document}

\title{Kronoseismology IV:\\ Six previously unidentified waves in Saturn's middle C ring}
\author{M. M. Hedman$^a$, P.D. Nicholson$^b$, R.G. French$^c$ }
\affil{$^a$ Department of Physics, University of Idaho, Moscow ID 83844-0903 \\
$^b$ Department of Astronomy, Cornell University, Ithaca NY 14853 \\
$^c$ Department of Astronomy, Wellesley College, Wellesley MA 02481 }
\begin{abstract}
Recent studies of stellar occultations observed by the Visual and Infrared Mapping Spectrometer (VIMS) onboard the Cassini spacecraft have demonstrated that multiple spiral wave structures in Saturn's rings are probably generated by normal-mode oscillations inside the planet. Wavelet-based analyses have been able to unambiguously determine both the number of spiral arms and the rotation rate of many of these patterns. However, there are many more planetary normal modes that should have resonances in the rings, implying that many normal modes do not have sufficiently large amplitudes to generate obvious ring waves. Fortunately, recent advances in wavelet analysis  allow weaker wave signals to be uncovered by combining data from multiple occultations. These new analytical tools reveal that a pattern previously identified as a single spiral wave actually consists of two superimposed waves, one with 5 spiral arms rotating at 1593.6$^\circ$/day  and one with 11 spiral arms rotating at 1450.5$^\circ$/day. Furthermore, a broad search for new waves revealed four previously unknown wave patterns with 6, 7, 8 and 9 spiral arms rotating around the planet at 1538.2$^\circ$/day, 1492.5$^\circ$/day, 1454.2$^\circ$/day and 1421.8$^\circ$/day, respectively. These six patterns provide precise frequencies for another six fundamental normal modes inside Saturn, {yielding what is now a complete sequence of fundamental sectoral normal modes with azimuthal wavenumbers from 2 to 10. These frequencies should place strong constraints on Saturn's interior structure and rotation rate, while} the relative amplitudes of these waves should help clarify how the corresponding normal modes are excited inside the planet.
\end{abstract}

\maketitle 

\section{Introduction}

Saturn's rings are an exquisitely sensitive dynamical system that can function as a seismometer for the planet. Scattered throughout the rings are tightly wound spiral patterns called density and bending  waves that are generated at locations where the orbital motions of the ring particles are in resonance with a periodic external force. Many of these features can be attributed to resonances with Saturn's various moons, but a growing number appear to be generated by asymmetries and/or oscillations within the planet itself, { confirming predictions made decades earlier \citep{Stevenson82, Marley90, Marley91, MarleyPorco93}. \citet{HN13} first used wavelet-based methods to determine the pattern speeds of six waves in the middle C ring that had 2, 3 and 4 spiral arms and appeared to be generated by either fundamental sectoral normal modes within the planet (i.e. modes with no radial nodes and $\ell=m$, where $\ell$ and $m$ are the standard indices for a spherical harmonic expansion) or mixtures of these fundamental modes with gravity modes \citep{Fuller14}. \citet{HN14} then applied the same basic techniques to seven additional waves, identifying one as a 10-armed spiral probably generated by another fundamental sectoral normal mode, along with a number of waves that appeared to be driven by persistent asymmetries in the planet's gravitational field rotating at roughly the planet's spin rate. Later, \citet{French16} adapted these techniques to characterize a wave within the eccentric Maxwell ringlet, and demonstrated that this wave was also probably generated by a $m=2$ fundamental sectoral normal mode. Finally, \citet{French18} examined a series of density and bending waves in the inner C ring and were able to determine the number of spiral arms and pattern speeds for six of them. Most of these also appear to be generated by fundamental normal modes, but these specific modes were most likely non-sectoral (i.e. they have $\ell \ne m$). All of these different planetary structures are providing new and novel insights into Saturn's internal structure  \citep[e.g.][]{Fuller14, M18}. }

\begin{figure}
\resizebox{3in}{!}{\includegraphics{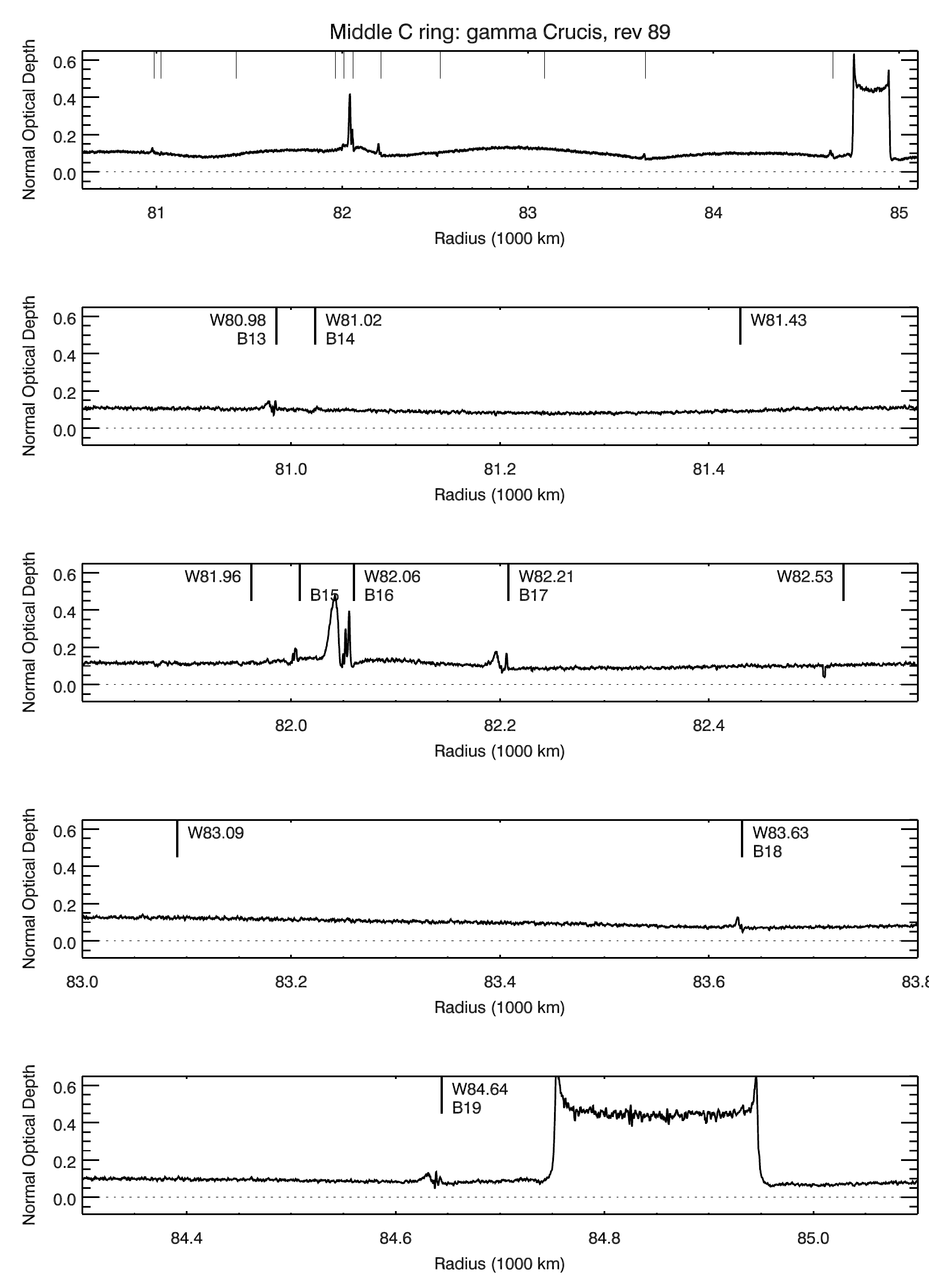}}
\caption{Overview of the portion of the middle C ring considered for this study. The top panel shows an optical depth profile of the region of interest derived from an occultation of the star $\gamma$ Crucis observed during Cassini Rev (i.e. orbit) 89, and the lower four panels show close ups of three portions of the above profile. Locations of potential wave signals are marked with lines, and are designated using two different notations \citep{Colwell09, Baillie11}. The features marked W80.98/B13, W82.00/B15, W82.06/B16, W82.21/B17, W83.63/B18 and W84.64/B19 are waves clearly visible in individual occultation profiles and were previously identified as being due to $m=4, 3,3,3,10$ and $2$ fundamental sectoral normal modes, respectively. Feature W81.02/B14 is a feature that was recognized by \citet{Baillie11}, but was not previously attributed to any specific planetary normal mode oscillations. Features W81.43, W81.96, W82.53 and W83.09 are wave candidates newly identified in this work.}
\label{ovfig}
\end{figure}

However, it is now clear that all the waves  identified in previous surveys \citep{Rosen91, Baillie11} represent only a fraction of the normal-mode oscillations inside the planet. For example, consider the waves that appear to be generated by fundamental sectoral normal modes. Thus far, the only waves generated by such modes that have been identified are those  with azimuthal wavenumbers  $m=2,3, 4$ and 10, most of which are found in a relatively bland part of the C ring between 80000 km and {84700 km} from Saturn's center (see Figure~\ref{ovfig}). However, modes with $m=5$ through $9$ should also generate resonances in this region, and there is only one previously recognized wave feature in this region whose identity has not yet been established (designated W81.02 in Figure~\ref{ovfig}). Hence there must be at least four planetary normal  modes that do not generate obvious density waves in the rings. Most likely, these normal mode oscillations are weaker, and so do not perturb the rings strongly enough to produce density waves with high enough amplitudes to be identified in previous surveys.

Fortunately, further improvements in wavelet-based analyses of the rings have enabled us to find and characterize much weaker density wave signatures than was previously possible. These techniques (which are described in more detail below) account for the expected phase shifts of the signals with a particular azimuthal wavenumber $m$ when co-adding wavelet transforms from multiple profiles. This is an extremely powerful filtering technique that has already enabled us to isolate signals from weak waves driven by satellites in Saturn's dense B ring \citep{HN16}. On the one hand, these methods have enabled us to determine that the previously-unidentified wave located around a radius of 81020 km is in fact  two waves that lie almost exactly on top of each other, one with five arms and the other with eleven arms. The 5-armed wave is probably generated by a resonance with the $m=5$ fundamental sectoral normal mode, while the 11-armed wave probably corresponds to a planetary normal mode with $\ell=13$ and $m=11$. On the other hand, these tools have revealed additional signatures that appear to be due to weak waves with 6, 7, 8 and 9 arms located near the expected resonances with fundamental sectoral normal modes { predicted by \citet{MarleyPorco93} and \citet{M18}}. These analyses therefore yield frequencies for six additional planetary normal mode oscillations, including the fundamental sectoral normal modes with $m$ values from 5 to 9, which should place strong constraints on the planet's internal structure. In addition, this work provides information about the relative amplitudes of these waves, which should help illuminate how these normal modes are excited inside the planet. 

Section~\ref{methods} below summarizes our analytical methods for finding and identifying these waves. Section~\ref{W81p02} then shows how these techniques can be used to untangle the two waves found around 81020 km, while Section~\ref{m6-9} describes how these techniques are used to find the weak waves generated by other sectoral normal modes. Section~\ref{wavelets} discusses detailed wavelet analyses of all these features, which yield estimates for both their pattern speeds and their amplitudes. Finally, Section~\ref{summary} summarizes our findings.

\bigskip

\section{Methods}
\label{methods}

\subsection{Observational Data}

This analysis focuses on stellar occultation data obtained by the Visual and Infrared Mapping Spectrometer (VIMS) onboard the Cassini Spacecraft \citep{Brown04}. While in its standard operating mode VIMS obtains spatially resolved spectra of various objects in the Saturn system, this instrument can also operate in a mode where it repeatedly measures the spectrum of a star as the planet or its rings pass between the star and the spacecraft. In this occultation mode, a precise time-stamp is appended to each spectrum to facilitate reconstruction of the observation geometry.  

As with previous occultation studies, here we will only consider data obtained at wavelengths between 2.87 and 3.00 microns, where the rings are especially dark and so provide a minimal background to the stellar signal. Using the appropriate SPICE kernels \citep{Acton96}, we use the timing information encoded with the occultation data to compute both the radius and inertial longitude in the rings that the starlight passed through. Note that the information encoded in these kernels has been determined to be accurate to within one kilometer, and fine-scale adjustments based on the positions of circular ring features enable these estimates to be refined to an accuracy of order 150 m. For this analysis, we use the latest estimates of these offsets from \citet{French17}; we have verified that these new offsets do not change the results described in \citet{HN14}

The VIMS instrument has a highly linear response function \citep{Brown04}, so the raw data numbers returned by the spacecraft are directly proportional to the apparent brightness of the star. We can therefore estimate the transmission through the rings $T$ as simply the ratio of the observed signal at a given radius to the average signal in regions outside the rings. From this transmission, we can compute the ring's optical depth  $\tau$ using the standard formula $\tau=-\ln(T)$. Both $T$ and $\tau$ depend on the observation geometry, but for relatively low optical depth regions like the middle C ring we can define the normal optical depth $\tau_n=\tau/|\sin(B)|$  ($B$ being the ring opening angle to the star), which should have nearly the same value for all the occultations considered here.

We will consider three different groups of occultations for this study:
\begin{itemize}\item[A.] For our initial investigation of the W81.02 patterns (Section~\ref{W81p02}), we use the same 23 occultation cuts obtained between 2007 and 2014 listed in Table 2 of \citet{HN14}. 
\item[B.] For our initial search for waves with $m=-6$ through $-9$ (Section~\ref{m6-9}), we consider the subset of occultations that use the star $\gamma$ Crucis and were obtained between Revs 71 and 102 (i.e. 2008-2009) listed in Table~\ref{obstab}. 
\item[C.] Finally, for our more in-depth analysis of the wave candidates (Section~\ref{wavelets}), we utilize a larger set of 56 occultations summarized in Table~\ref{obstab}. This set includes all occultations obtained prior to Rev (i.e. Cassini orbit) 270 that cover the relevant wave features and satisfy the following requirements:
\begin{itemize}
\item Do not have any data gaps larger than 1 km in the region of interest for the wave (see Table~\ref{wavetab}).
\item Have mean normal optical depth values within the region of interest (see Table~\ref{wavetab}) that are within 0.1 of the median value for all occultations. This eliminates occultations which have unstable signal levels and other instrumental issues that could impact the relevant signals.
\item Have $rms$ normal optical depth variations smaller than 0.015 on radius scales of 0.1 km in nearby featureless ring regions (80700-80800 km, 81200-81300 km,  82300-82350 km, 83700-83800 km or 84100-84200 km). This removes occultations with low signal-to-noise.
\end{itemize}
\end{itemize}

\begin{table*}
\caption{Occultations used for this study}
\label{obstab}
\hspace{-1.5in}{\resizebox{8in}{!}{\begin{tabular}{|c|c|c|c|c|c| c|c|c|c|c|c|c|c|c|c|c|c|} \hline
Star & Rev & $^a$ & Date & B($^0$)$^b$ & $\lambda(^0)^c$ & W80.98 & W81.02a & W81.02b & W81.43 & W81.96 & W82.00 & W82.06 & W82.21 & W82.53 & W83.09 & W83.63 & W84.64 \\ \hline
             RHya & 036 & i & 2007-001 & -29.4 & 173.1-177.7 & X & X & X & X & X & X & X & X & X & X & X & X \\
   $\alpha$Aur & 041 & i & 2007-082 & +50.9 & 342.5-348.4 & X & X & X & X & X & X & X & X &  X & X & X & X \\
$\gamma$Cru & 073 & i & 2008-174 & -62.3 & 182.0-182.7 & X & X & X & X & X & X & X & X & X & X & X & X \\
$\gamma$Cru & 078 & i & 2008-209 & -62.3 & 180.7-181.4 & X & X & X & X & X & X & X & X & X & X & X & X \\
$\gamma$Cru & 079 & i & 2008-216 & -62.3 & 179.0-179.9 & X & X & X & X & X & X & X & X & X & X & X & X \\
           RSCnc & 080 & i & 2008-226 & +30.0 & 82.4-93.0    & X & X & X & X & X & X & X & X & X & X & X & X \\
           RSCnc & 080 & e& 2008-226 & +30.0 & 118.7-129.4 & X & X & X & X & X & X & X & X & X & X & X & X \\
$\gamma$Cru & 081 & i & 2008-231 & -62.3 & 178.1-179.0 & X & X & X & X & X & X & X & X & X & X & X & X \\
$\gamma$Cru & 082 & i & 2008-238 & -62.3 & 177.7-178.6 & X & X & X & X & X & X & X & X & X & X & X & X \\
            RSCnc  & 85 & i & 2008-263 & +30.0 & 88.3-97.7    &     &     &    &     &    &  X & X & X & X & X & X & X \\
            RSCnc  & 85 & e& 2008-263 & +30.0 & 113.6-123.1 &    &     &    &     &    & X & X & X & X & X & X & X \\
$\gamma$Cru & 086 & i & 2008-268 & -62.3 & 176.6-177.6 & X & X & X & X & X & X & X & X & X & X & X & X \\
            RSCnc  & 87 & i & 2008-277 & +30.0 & 91.7- 99.3    &    &     &    &     &    &    &     &     &    & X & X & X \\
            RSCnc  & 87 & e& 2008-277 & +30.0 & 117.9-119.6 &    &     &    &     &    &    &     &     &    & X & X & X \\
$\gamma$Cru & 089 & i & 2008-290 & -62.3 & 176.4-177.4 & X & X & X & X & X & X & X & X & X & X & X & X \\
$\gamma$Cru & 093 & i & 2009-320 & -62.3 & 207.5-208.4 & X & X & X & X & X & X & X & X & X & X & X & X \\
$\gamma$Cru & 094 & i & 2008-328 & -62.3 & 191.7-191.7 & X & X & X & X & X & X & X & X & X & X & X & X \\
$\gamma$Cru & 100 & i & 2009-012 & -62.3 & 222.6-224.7 & X & X & X & X & X & X & X & X & X & X & X & X \\
$\gamma$Cru & 102 & i & 2009-031 & -62.3 & 222.3-224.4 & X & X & X & X & X & X & X & X & X & X & X & X \\
     $\beta$Peg & 104 & i & 2009-057 &+31.7 & 342.1-344.3 & X & X & X & X &    & X &     & X & X & X & X & X \\
               RCas & 106 & i & 2009-082 & +56.0 & 79.5- 90.6  &   & X & X & X & X & X & X & X & X & X & X & X \\
   $\alpha$Sco & 115 & i & 2009-209 & -32.2 & 157.4-159.8 & X & X & X & X & X & X & X & X & X & X & X & X \\
     $\beta$Peg & 170 & e& 2012-224 & +31.7 & 78.1-79.8     & X & X & X & X & X & X & X & X & X & X & X & X \\
     $\beta$Peg & 172 & i & 2012-266 & +31.7 & 310.9-312.7 & X & X & X & X & X & X & X & X & X & X & X & X \\
 $\lambda$Vel & 173 & i & 2012-292 & -43.8 & 148.4-152.8 & X & X & X & X & X & X &     &     & X & X & X & X \\
             WHya & 179 & i & 2013-019 & -34.6 & 143.9-147.2 & X & X & X & X & X & X & X & X & X & X & X & X \\
             WHya & 180 & i & 2013-033 & -34.6 & 144.4-147.8 & X & X & X & X & X & X & X & X & X & X & X & X \\
             WHya & 181 & i & 2013-049 & -34.6 & 144.4-147.8 & X & X & X & X & X & X & X & X & X & X & X & X \\
      $\mu$Cep & 185 & e& 2013-090 & +59.9 & 43.1-48.6    & X & X & X & X & X & X & X & X & X & X & X & X \\
             WHya & 186 & e& 2013-103 & -34.6 & 297.5-298.6 & X & X & X & X & X & X & X & X & X & X & X & X \\
$\gamma$Cru & 187 & i & 2013-112 & -62.3 &148.3-152.8 & X & X & X & X & X & X & X & X & X & X & X & X \\
$\gamma$Cru & 187 & e& 2013-112 & -62.3 & 224.5-228.9 & X &  X & X & X & X & X & X & X & X & X & X & X \\
             WHya & 189 & e& 2013-132 & -34.6 & 296.3-297.4 & X & X & X & X & X & X & X & X & X & X & X & X \\
      $\mu$Cep & 191 & i & 2013-148 & +59.9 & 289.7-290.4 & X &   & X & X & X & X & X & X & X & X & X & X \\
      $\mu$Cep & 193 & i & 2013-172 & +59.9 & 289.7-290.4 & X & X & X &X & X & X & X & X & X & X & X & X \\
              2Cen & 194 & i & 2013-189 & -40.7 & 150.6-155.0 &     &    &    & X & X &     &    &    & X & X & X & X \\
              2Cen & 194 & e& 2013-189 & -40.7 & 227.7-232.0 & X & X & X & X & X &    &    &     & X &    & X   &   \\
               RLyr & 198 & i & 2013-289 & +40.8 & 260.9-262.1 & X & X & X & X & X &     & X & X & X & X & X & X \\
               RLyr & 199 & i & 2013-337 & +40.8 & 227.6-231.9 & X & X & X & X & X & X & X & X & X & X & X & X \\
               RLyr & 200 & i & 2014-003 & +40.8 & 255.7-257.1 & X &  X & X & X & X & X & X & X & X & X & X & X \\
             L2Pup & 201 & i & 2014-051 & -41.9 & 95.2-95.4     & X & X & X & X & X & X &    & X & X & X & X & X \\
               RLyr & 202 & e& 2014-067 & +40.8 & 54.3-57.3     & X & X & X & X & X & X & X & X & X & X & X & X \\
            L2Pup & 205 & e& 2014-175 & -41.9 & 219.0-224.0 &  X & X & X & X & X & X & X & X & X & X & X & X \\
               RLyr & 208 & e& 2014-262 & +40.8 & 44.3-47.6     &     & X & X & X & X & X & X & X & X &    &   & X \\
   $\alpha$Sco & 241 & e& 2016-243 & -32.2 & 21.9-27.3      & X & X & X & X & X & X & X & X & X & X & X & X \\
   $\alpha$Sco & 243 & e& 2016-267 & -32.2 & 20.7-25.9      & X & X & X & X & X & X & X & X & X & X & X & X \\
   $\alpha$Sco & 245 & e& 2016-287 & -32.2 & 19.4-25.1      & X & X & X & X & X &    & X & X & X & X & X & X \\
$\gamma$Cru & 245 & e& 2016-286 & -62.3 & 257.2-269.4 & X & X & X & X & X & X & X & X & X & X & X & X \\ 
$\gamma$Cru & 255 & i & 2017-001 & -62.3 & 146.8-147.1 &     & X & X & X & X & X & X & X & X & X & X & X \\
$\gamma$Cru & 264 & i & 2017-086 & -62.3 & 145.0-145.2 & X & X & X & X & X & X & X & X & X & X & X & X \\
$\lambda$Vel & 268 & i & 2017-094 & -43.8 & 134.4-135.3 &     & X & X & X & X &     & X &     & X & X & X &    \\
$\gamma$Cru & 268 & i & 2017-095 & -62.3 & 143.9-144.2 & X & X &  X & X & X & X & X & X & X & X & X & X \\
     $\alpha$Ori & 268 & e& 2017-096 & +11.7 & 196.4-201.0 & X & X & X & X & X & X & X & X & X & X & X & X \\
           VYCMa & 269 & i & 2017-100 & -23.4 & 201.5-206.8 & X & X & X & X & X & X & X & X & X & X & X & X \\
$\gamma$Cru & 269 & i & 2017-102 & -62.3 & 143.8-144.1 & X & X & X & X & X & X & X & X & X & X & X & X \\
     $\alpha$Ori & 269 & e& 2017-104 & +11.7 & 6.1-9.6       & X & X & X & X & X & X & X & X & X & X & X & X \\
\hline
           \end{tabular}}}

$^a$ i=ingress occultation, e= egress occultation

$^b$ Ring opening angle to star (positive indicates star is north of Saturn's equatorial plane)

$^c$ Span of inertial longitudes, measured relative to the ascending node of the ring particles on the J2000 coordinate system.          
           
\end{table*}

\subsection{Analytical approach}

For the purposes of this study, the two most important parameters associated with these waves are the number of spiral arms $|m|$ and the pattern speed $\Omega_p$ at which these density variations rotate around the planet in inertial space \citep{Shu84}. For waves generated by planetary normal-mode oscillations, the number of spiral arms equals the mode's azimuthal wavenumber, while the pattern speed equals the mode's propagation rate around the planet in inertial space. In addition, a wave with a given number of spiral arms and pattern speed can only be generated at resonant locations $r_L$ where the ring-particles' orbital mean motion $n_L$ and radial epicyclic frequency $\kappa_L$ satisfy the following relationship:
\begin{equation}
m(n_L-\Omega_p)=\kappa_L.
\label{pateq}
\end{equation} 

Note that in this expression we allow $m$ to be a signed quantity, such that $m>0$ corresponds to cases where the pattern speed is slower than the mean motion (i.e Inner Lindblad Resonances), and $m<0$ corresponds to cases where the pattern speed is faster than the mean motion (i.e. Outer Lindblad Resonances). If a wave is observed at a given radius, the corresponding orbital frequencies $n_L$ and $\kappa_L$ can be determined from the planet's gravitational field, hence there is a discrete set of pattern speeds the wave could have, one for each possible value of $m$.

Fortunately, both $m$ and $\Omega_p$ can be estimated by comparing wave profiles observed at different times and longitudes.  A generic density wave with $|m|$ arms and pattern speed $\Omega_p$  causes the surface mass density of the ring $\sigma$ to vary with radius $r$, longitude $\lambda$ and time $t$ as follows:
\begin{equation}
\sigma=\sigma_0\Re\left[1+A(r)e^{i[\phi_r(r)+|m|(\lambda-\Omega_p t)+\phi_0]}\right],
\label{wave}
\end{equation}
where $\sigma_0$ and $\phi_0$ are constants, $A(r)$ is a slowly-varying function of radius, and $\phi_r(r)$ is the radius-dependent part of the wave's phase, which has the following form at sufficiently large distances from the resonant radius $r_L$ (so long as $m \ne 1$, Shu 1984):
\begin{equation}
\phi_r(r)=\frac{3|m-1| M_P(r-r_L)^2}{4\pi\sigma_0 r_L^4},
\label{phir}
\end{equation}
where $M_P$ is the planet's mass. Note that waves with $m>0$ can only propagate exterior to the resonant radius, while waves with $m<0$ can only propagate interior to the resonant radius.  

For the entire region of interest here, the ring's optical depth appears to be directly proportional to its surface mass density \citep{Hedman11}, so in any given occultation the ring's optical depth will  vary quasi-sinusoidally, but the positions of the peaks and troughs will vary with longitude and time in a matter that is sensitive to $m$ and $\Omega_p$. Indeed, wavelet-based methods allow us to quantify the phase differences between optical depth profiles, and thereby determine the values of $m$ and $\Omega_p$ for a large number of waves that appear to be generated by structures inside the planet \citep{HN13, HN14, French16, French18}.

Our previous analyses of the planet-generated waves employed wavelet-based algorithms which estimated the average phase differences between pairs of wave profiles and then compared those differences with those expected given different assumptions about $m$ and $\Omega_p$ \citep{HN13, HN14, French16, French18}. While powerful, these algorithms have two important limitations: they assume there is a single wave in the analyzed region of interest, and they assume the wave signal dominates the optical depth variations in individual profiles. Here we will relax these assumptions by adapting a different set of algorithms to isolate waves with particular pattern speeds and number of spiral arms within the occultation data.  The details of this approach are described in \citet{HN16}, but for the sake of completeness we will summarize the important aspects of the procedures here.

We begin by taking each occultation profile and interpolating the transmission values onto a regular grid of radii with a spacing of 100 meters, converting the profile to normal optical depth\footnote{Our previous wavelet-based analyses performed wavelet transformations on the transmission profiles. However, since in this case the relative amplitudes of the waves are important, it makes sense to perform the wavelet analysis on the quantity that is independent of viewing geometry. We have verified that we obtain the same pattern speeds if we use either the transmission or the optical depth.}, and then transforming the profile into a wavelet  using the IDL  {\tt wavelet} routine  \citep{TC98} with a Morlet mother wavelet and parameter $\omega_0=6$. This yields a complex wavelet for each profile $\mathcal{W}_i=\mathcal{A}_ie^{i\Phi_i}$ where $\mathcal{W}_i, \mathcal{A}_i$ and $\Phi_i$ are all functions of both radius $r$ and wavenumber $k$. For the signal from a density wave the wavelet phase $\Phi_i$ is equivalent to the wave phase in Equation~\ref{wave}. Hence, given the observed longitude $\lambda_i$ and observation time $t_i$ for each occultation, we can compute the phase parameter $\phi_i=|m|[\lambda_i-\Omega_p t_i]$ for any chosen values of  $|m|$ and $\Omega_p$. For a wave with the selected parameters, the phase difference $\Phi_i-\phi_i =\phi_r(r)+\phi_0$ for every occultation, so we can define a phase corrected wavelet:
\begin{equation}
\mathcal{W}_{\phi,i}=\mathcal{W}_ie^{-i\phi_i}=\mathcal{A}_ie^{i(\Phi_i-\phi_i)}.
\end{equation} 
For signals with the selected $m$ and $\Omega_p$, the phase will be the same for all the occultation profiles, so the average phase corrected wavelet
\begin{equation}
\langle \mathcal{W}_\phi \rangle=\frac{1}{N}\sum_{i=1}^N\mathcal{W}_{\phi,i}
\end{equation}
will be nonzero, while any signal without these properties will average to zero. Thus only the desired signal should remain in the power of the average phase corrected wavelet
\begin{equation}
\mathcal{P}_\phi(r,k)=|\langle W_\phi \rangle|^2=\left|\frac{1}{N}\sum_{i=1}^N\mathcal{W}_{\phi,i}\right|^2,
\end{equation}
while all other signals are only seen in the average wavelet power:
\begin{equation}
\bar{\mathcal{P}}(r,k)=\langle |W_\phi|^2 \rangle=\frac{1}{N}\sum_{i=1}^N\left|\mathcal{W}_{\phi,i}\right|^2.
\end{equation}
Hence the ratio of these two powers $\mathcal{R}(r,k)=\mathcal{P}_\phi/\bar{\mathcal{P}}$ \citep[which ranges between 0 and 1, see][]{HN16} provides a measure of how much of the signal is consistent with the assumed $m$  and $\Omega_p$.

The average phase-corrected wavelet can also be used to produce a reconstructed profile of the part of the signal with the selected $m$-number and pattern speed. This is accomplished by taking the average phase-corrected wavelet and applying the inverse wavelet transformation. In practice, we only consider a finite range of wavenumbers when performing this inversion to remove residual high-frequency noise and slow background trends. The resulting profile is complex, but the real and imaginary parts of the profile just correspond to absolute wave phases of 0 and $\pi/2$, respectively. Since the wavelets were computed from the normal optical depth profiles, this reconstructed profile gives the normal optical depth variations associated with the wave. We divide these variations by the average optical depth profile to create a profile of the fractional optical depth variations. These fractional optical depth variations are easier to compare among different waves. We will therefore plot the real part of these optical depth variations, and report the peak amplitude of these variations as the maximum value of the square root of the sum of the squares of the real and imaginary parts of the profile.

\begin{figure}[tbhp]
\resizebox{3.05in}{!}{\includegraphics{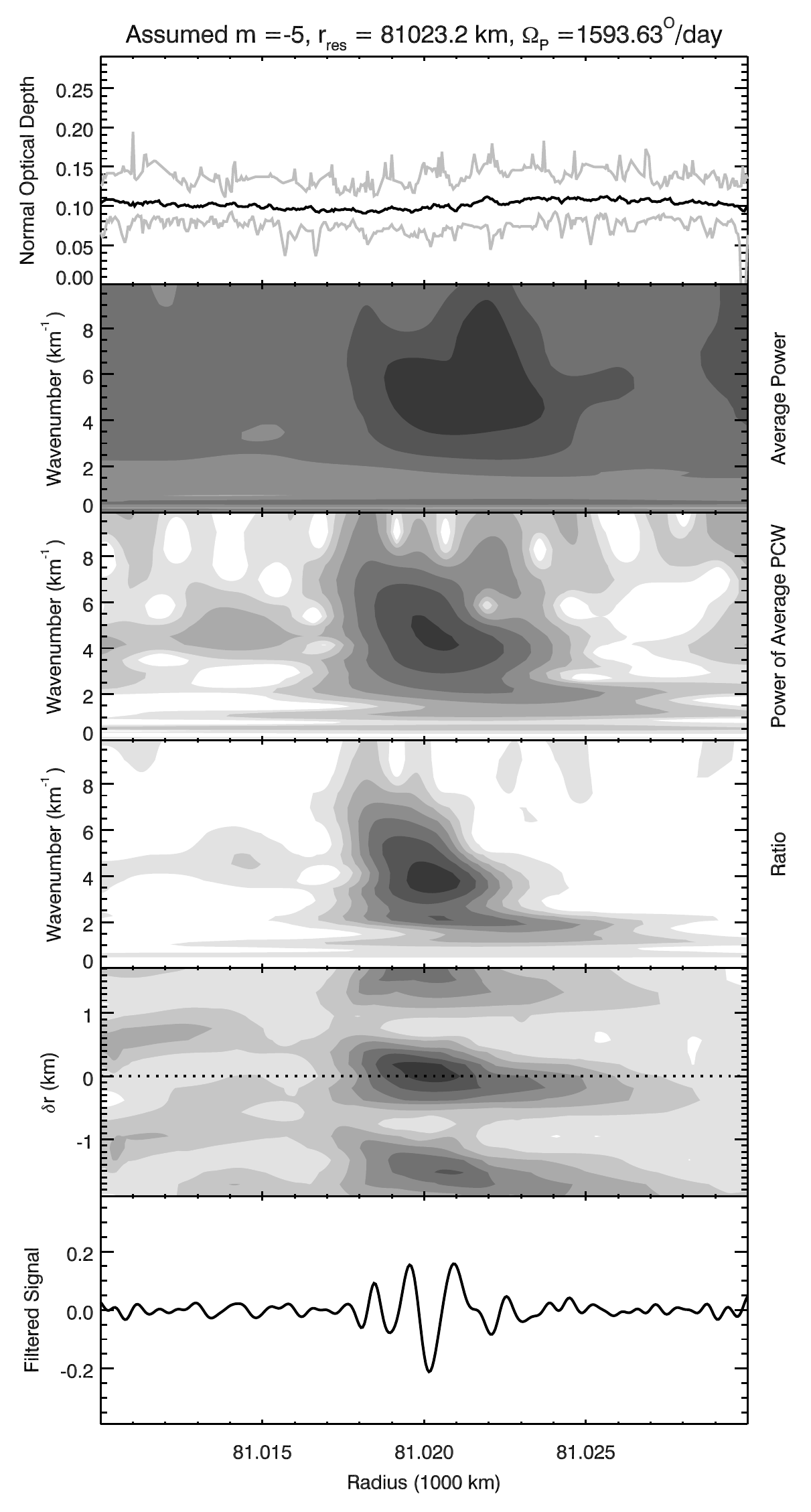}}
\caption{Phase-corrected wavelet analysis of W81.02 wave. The top panel shows the average optical depth of the ring near this wave as a black line, and the range of normal optical depths observed among the occultations is shown with the grey lines. The next panel shows the average wavelet power as a function of wavenumber and radius, while the  third panel shows the power of the average phase-corrected wavelet assuming $m=-5$ and $\Omega_p=1593.63^\circ$/day. The fourth panel shows the ratio of these two powers, which is a measure of how much of the signal is consistent with the selected values of $m$ and $\Omega_p$. In both these panels there is a  clear signal centered around 81020 km. The fifth panel shows the peak power ratio between wavelengths of 1 and 3 km as a function of radius and pattern speed (expressed as a shift in the nominal resonant radius of 81023.3 km). The bottom panel shows the reconstructed wave profile derived from the average phase-corrected wavelet, considering only wavelengths between 0.5 and 2.0 km, which looks like a sensible inward-propagating density wave, as expected for this resonance.}
\label{m5wavesep}
\end{figure}

\begin{figure}[tbh]
\resizebox{3.05in}{!}{\includegraphics{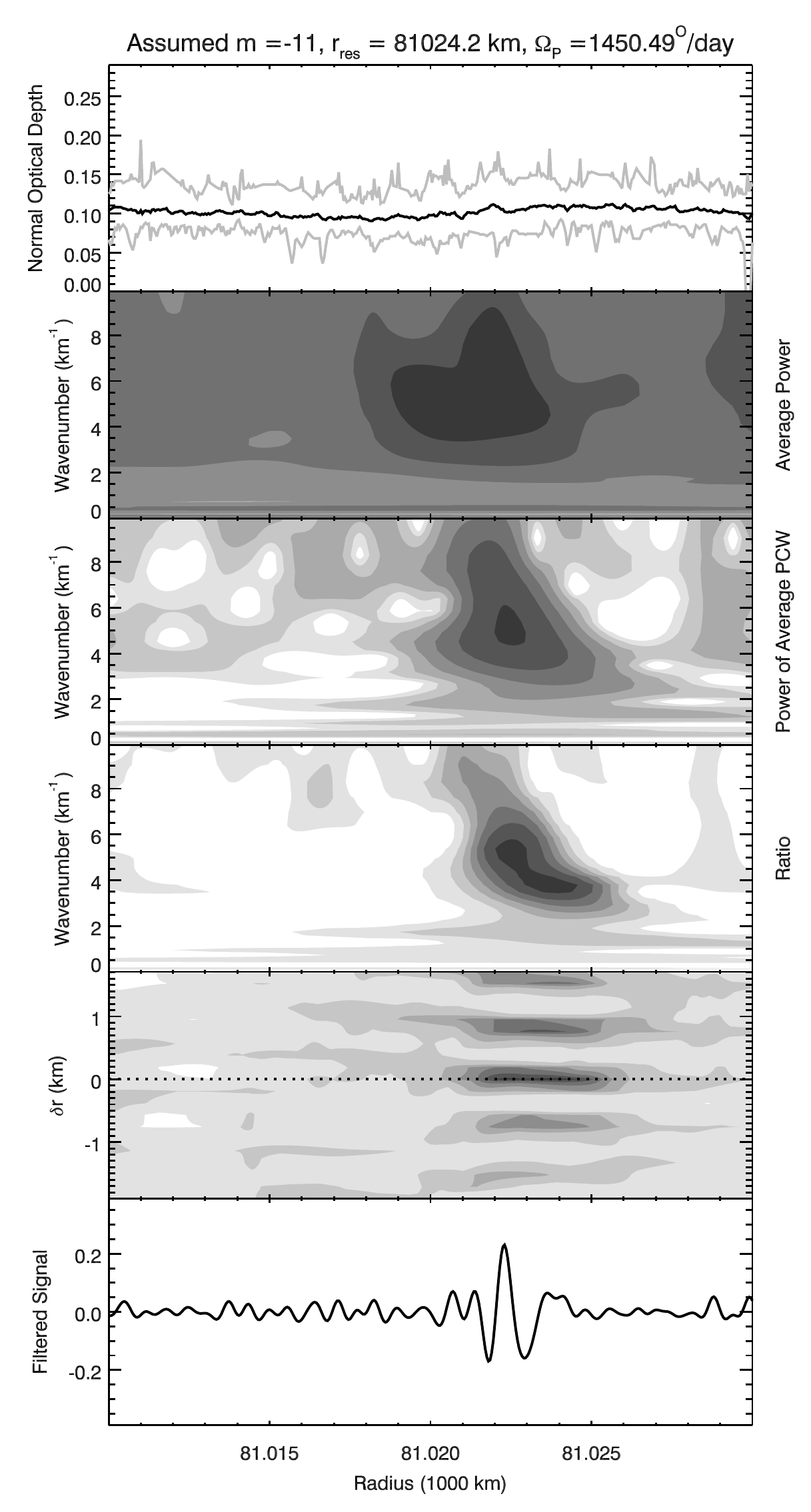}}
\caption{Phase-corrected wavelet analysis of W81.02 wave, with the same format as Figure~\ref{m5wavesep}, except that here we assume $m=-11$, $\Omega_p=1450.49^\circ$/day and a nominal resonant radius of 81024.2 km. Note that in this case, the peak signal falls around 81023 km, and the reconstructed wave profile is also centered at this location.}
\label{m11wavesep}
\end{figure}

\section{Untangling the W81.02 waves}
\label{W81p02}

Figures~\ref{m5wavesep} and ~\ref{m11wavesep} show the results of the phase-corrected wavelet analysis of W81.02, using the Group A observations \citep[i.e. the same ones used in][]{HN14}. In each plot, we show the average wavelet power, the power of the average phase-corrected wavelet, and the ratio of these two powers as functions of wavenumber and location, along with the peak power ratio as functions of radius and pattern speed (expressed as a shift in the nominal resonance position). The two plots shown here are for $m=-5$ and $m=-11$, the two values for which \citet{HN14} found potential signals (no other $m$-numbers show such obvious patterns).

 In both cases, there is a clear signal in power ratio indicating that there is a pattern with the expected $m$-number and pattern speed. Both show a trend where the wavenumber decreases with increasing radius, consistent with inward-propagating waves. However, the critical aspect of these plots is that the signals in the two plots occur at slightly different locations and have correspondingly different resonant radii. For $m=-5$, the peak signal occurs between 81018 and 81021 km, and has a pattern speed corresponding to a resonant radius of 81023.2 km. By contrast, the $m=-11$ signal has its peak signal between 81021 and 81024 km, and a resonant radius about one kilometer further out, around 81024.2 km. Furthermore, by inverting the wavelet transformation, we can reconstruct the portion of the signal with the selected wavenumbers and pattern speed. These profiles are shown in the bottom panels of Figures~\ref{m5wavesep} and~\ref{m11wavesep}, and both look like sensible inward-propagating density waves. We therefore conclude  that W81.02 is indeed made up of two different waves that are partially overlapping each other. We will here designate the waves with 5 and 11 spiral arms as W81.02a and W81.02b, respectively.

\begin{figure}
\resizebox{3.in}{!}{\includegraphics{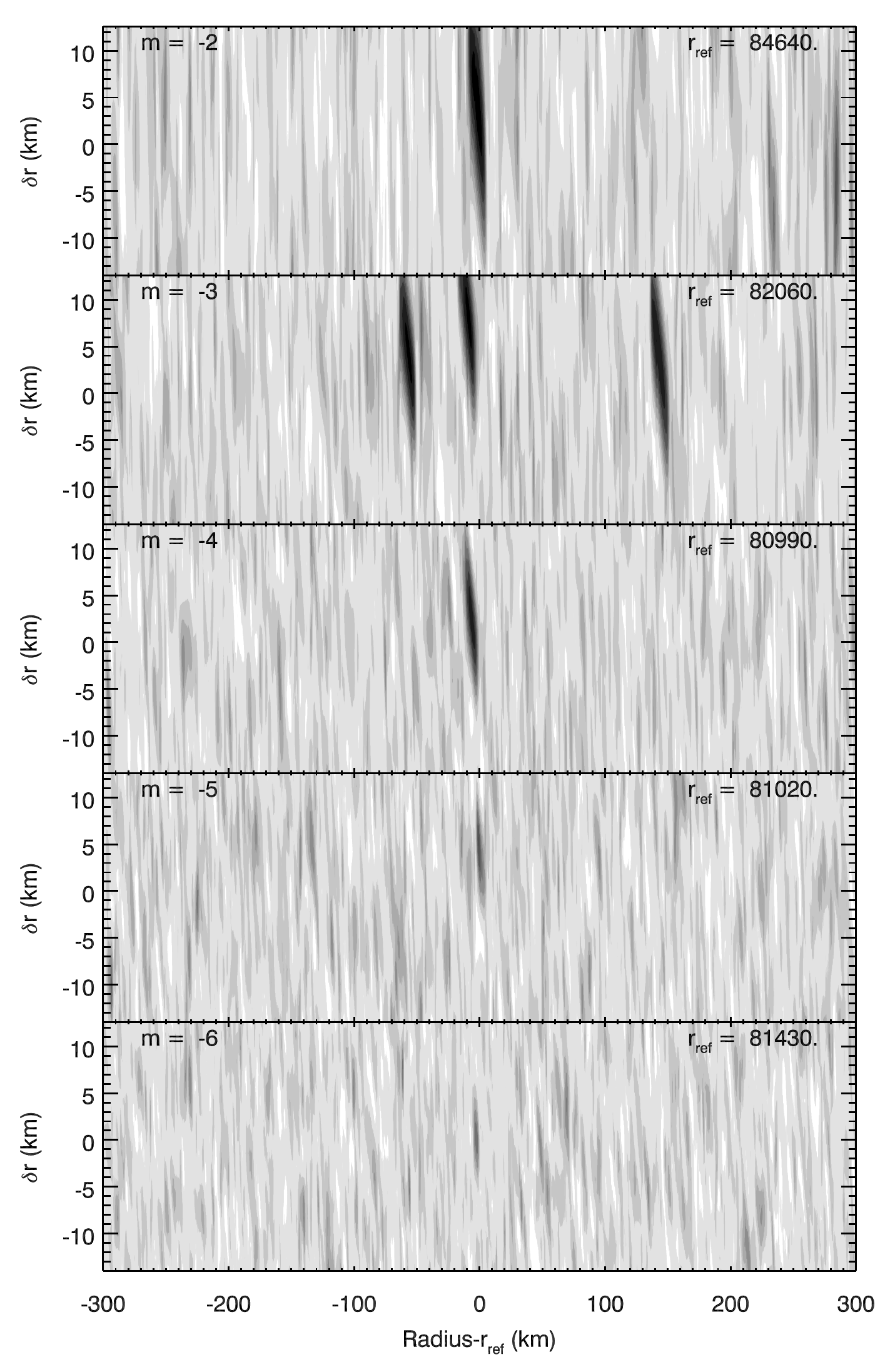}}
\caption{Search results for waves with $m=-2$ through $m=-6$ based on the Group B  $\gamma$ Crucis occultations. Each panel shows the peak wavelet power ratio for wavelengths between 1 and 3 km as a function of location in the rings and assumed pattern speed for the indicated $m$. In these plots, the nominal pattern speed (corresponding to $\delta r=0$) varies with position such that the assumed resonant radius equals the observed  radius. For inward-propagating waves, the strongest signals should have a pattern slightly slower than this, and so the signals should appear at slightly positive values of $\delta r$. Such signals are clearly seen for $m=-2,-3,-4$, corresponding to previously-known waves. The $m=-5$ signal also occurs at the location of the W81.02 wave. Finally, there is a weak but distinct $m=-6$ signal near 81430 km.}
\label{m2-6search}
\end{figure}

\begin{figure}
\resizebox{3.in}{!}{\includegraphics{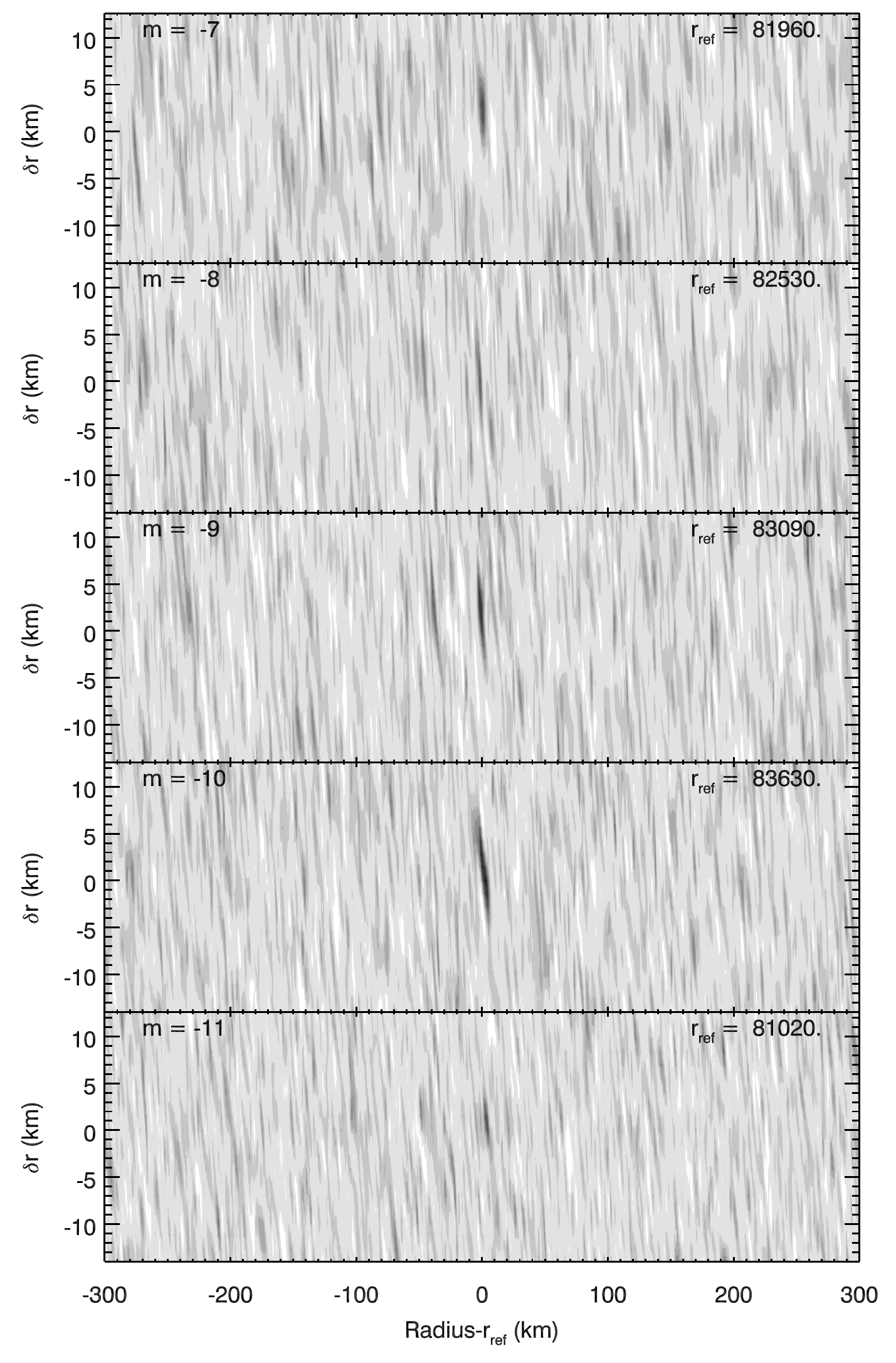}}
\caption{Search results for waves with $m=-7$ through $m=-11$  based on the Group B $\gamma$ Crucis occultations. Here the signal associated with the previously-identified $m=-10$ wave W83.63 is clear, and an $m=-11$ signal at W81.02 can also be identified. Furthermore, there are reasonably strong and clear $m=-7$ and $m=-9$ wave signals at 81960 km and 83090 km, respectively. A weaker signal for $m=-8$ can also be seen around 82530 km.}
\label{m7-11search}
\end{figure}

Neither of the two waves within W81.02 can be attributed to any known resonance with any of Saturn's moons, and so both are most likely generated by normal mode oscillations inside the planet itself. The five-armed spiral wave W81.02a falls just outside the $m=-4$ density wave W80.98, which we had previously identified as likely generated by the $m=4$ planetary sectoral normal mode \citep{HN13}. \citet{MarleyPorco93} predicted that the $m=5$ fundamental sectoral normal mode would generate a resonance in the middle C r-ring within 50 km of the $m=4$ sectoral normal mode, and so W81.02a is mostly likely generated by  the $m=5$ fundamental sectoral normal mode inside the planet. 

By contrast, W81.02b is probably not generated by a fundamental sectoral normal mode because it falls well interior to W83.63, a wave that was previously identified as $m=-10$ and is likely generated by the $m=10$ fundamental sectoral normal mode \citep{HN14}. Instead, W81.02b is likely generated by a normal mode with $m=11$ and $\ell=13$. The resonance location of this mode was not explicitly calculated by \citet{MarleyPorco93} or \citet{Marley14}, but extrapolating trends from their calculations of other modes with $\ell=m+2$ does place the wave at approximately the right location, and this is consistent with recent analyses by \citet{M18}.





\section{Searching for additional waves}
\label{m6-9}

Average phase-corrected wavelets can also be used to search for weak waves that cannot be identified in individual occultation profiles. In principle, one can process the occultations for all possible values of $m$ and $\Omega_p$ and search for large values of the power ratio $\mathcal{R}$ that could be indicative of a density wave. In practice, a wave with $|m|$ arms  near a given radius $r$ should have a pattern speed close to that given by Equation~\ref{pateq}, greatly reducing the parameter space that needs to be searched. 

There are many potential resonances with planetary normal modes that could produce weak waves in the rings. However, for this initial search we will focus on the waves generated by fundamental sectoral modes with $m$ between $-5$ and $-10$, which should all lie between the previously-identified W81.02a and W83.63 waves shown in Figure~\ref{ovfig} \citep{MarleyPorco93, Marley14, M18}. These waves should therefore all fall between 81000 km and 83500 km, which is a rather bland region without many structures that could interfere with the weak wave signals (see Figure~\ref{ovfig}). 

In principle, we could search for wave signals using average phase-corrected wavelet of all the profiles listed in Table~\ref{obstab}. However, in practice this is not an ideal approach because these data span such a long time interval that the strength of the signal would be extremely sensitive to small changes in the pattern speed. The pattern speed would therefore need to be sampled extremely finely to avoid missing the desired signals, which is inefficient and computationally expensive. Hence for this search we  only considered the $\gamma$ Crucis occultations obtained between Revs 71 and 102 (i.e. Group B in Section~\ref{methods}). These form a set of high-quality occultations from a relatively short time period. We can therefore sample pattern speed space more coarsely and not worry about missing important signals.

Figures~\ref{m2-6search} and~\ref{m7-11search} show the results of this search. Each panel of these figures shows the peak value of the power ratio $\mathcal{R}$ between wavelengths of 1 km and 3 km as a function of position and pattern speed, expressed as an offset from an assumed resonant radius. Unlike Figures~\ref{m5wavesep} and~\ref{m11wavesep}, where the assumed resonant radius is a single number, here the resonant radius varies such that it always equals the observed radius. Thus the signal  for $m=-2$ at $\delta r = 5$ km at a radius of 84640 km indicates that there is a signal at 84640 km with a pattern speed appropriate for $m=-2$ and a resonant radius around 84645 km.  Note the wavelength range considered here corresponds to the typical wavelengths of the previously identified waves in this region. Including longer wavelengths made the results more sensitive to slow drifts and offsets in the baseline signal levels, while including shorter wavelengths introduced more noise into the images.

The previously identified waves W84.64 ($m=-2$), W82.00, W82.06, W82.21 ($m=-3$), W80.98 ($m=-4$) and W83.63 ($m=-10)$ all appear as obvious dark patches in the corresponding panels with the appropriate $m$-values. The two components of W81.02 can also be seen in the $m=-5$ and $m=-11$ panels as small dark spots. Note that for all these signals the peak signal falls at a slightly positive values for $\delta r$, which makes sense because these are all inward-propagating waves, so the true resonance location falls slightly outside the location of the wave itself. The magnitude of this offset varies because the peak signal occurs where the wavelength of the wave is around 2 km, which can occur at different distances from the resonance depending on the number of arms in the wave and the local surface mass density. 

Small dark spots are also visible in the  $m=-6,-7,-8$ and $-9$ panels  at around 81430 km, 81960 km, 82530 km and 83090 km, respectively. Figures~\ref{m2-6search} and~\ref{m7-11search} show that each of these locations yields the strongest wavelet power ratio signature within 300 km, providing some evidence that these signals are real structures and not simply random noise. Furthermore, it is worth noting that the $m=-7$ and $m=-9$ signals at 81960 km and 83090 km look to be about as strong as $m=-5$ and $m=-11$ waves, while the $m=-6$ and $m=-8$ signals at 81430 km and 82530 km are noticeably weaker. This suggests that these four waves likely have a range of amplitudes, which turns out to be the case. For the sake of convenience, we will
follow the notation used by \citet{Colwell09}, and designate these new wave candidates W81.43, W81.96, W82.53 and W83.09.

\section{Detailed analysis of the wave signatures}
\label{wavelets}

Further evidence that the above signals do in fact represent real density waves can be obtained using the full suite of occultations listed in Table~\ref{obstab} (i.e. Group C of Section~\ref{methods}). This large number of occultations provides the strongest possible filter for signals with a particular $m$-number and pattern speed. Also, with occultations spanning an entire decade, the phase corrections become extremely sensitive to the assumed pattern speed, enabling very precise measurements of this parameter

Below we will consider the waves in order of decreasing signal strength. First we will review the properties of the previously-identified waves and show how they all have clear signatures with well-defined pattern speeds that can be used to produce sensible wave reconstructions. Next, we will consider the two components of W81.02 and derive refined estimates of their pattern speeds. Then we will consider the signatures with $m=-7$ and $m=-9$, which are the stronger and more robust of the newly discovered wave signatures. Finally, we will discuss the weak $m=-6$ and $m=-8$ signals and evaluate whether they represent real waves.

\subsection{Previously identified waves with $m=-2,-3,-4$ and $-10$}

\begin{figure}
\resizebox{3.in}{!}
{\includegraphics{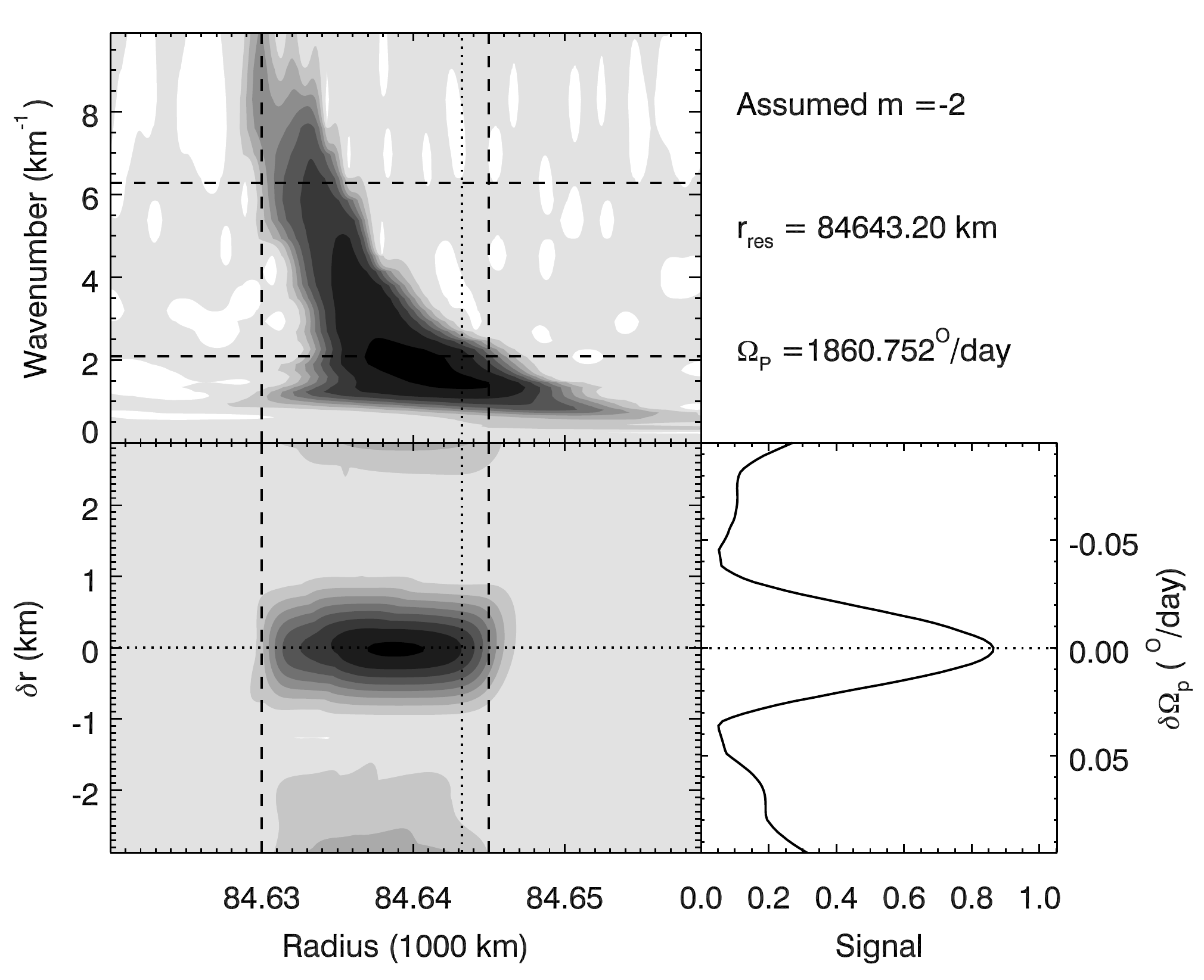}}
\caption{Detailed wavelet analysis of the $m=-2$ wave W84.64. The top panel shows the power ratio $\mathcal{R}$ as a function of wavenumber and radius assuming the indicated $m$-value and pattern speed. The vertical dotted line marks the location of the corresponding resonant radius, and the two horizontal dashed lines mark wavenumbers corresponding to wavelengths of 1 and 3 km. The bottom left panel shows the peak power ratio in the  wavelength range between 1 and 3 km as functions of radius and pattern speed (expressed as offsets from the expected pattern speed and resonant radius). The horizontal dotted line corresponds to the assumed pattern speed used in the upper panel. The lower right panel shows the peak power ratio versus pattern speed in the radial range marked with the vertical dashed lines in the other two panels. Note that the assumed pattern speed corresponds to the peak signal, and the wavelet power ratio shown in the top panel has the expected trend for an inward-propagating density wave (wavenumber increases with decreasing radius).}
\label{m2wave}
\end{figure}

\begin{figure}
\resizebox{3.in}{!}
{\includegraphics{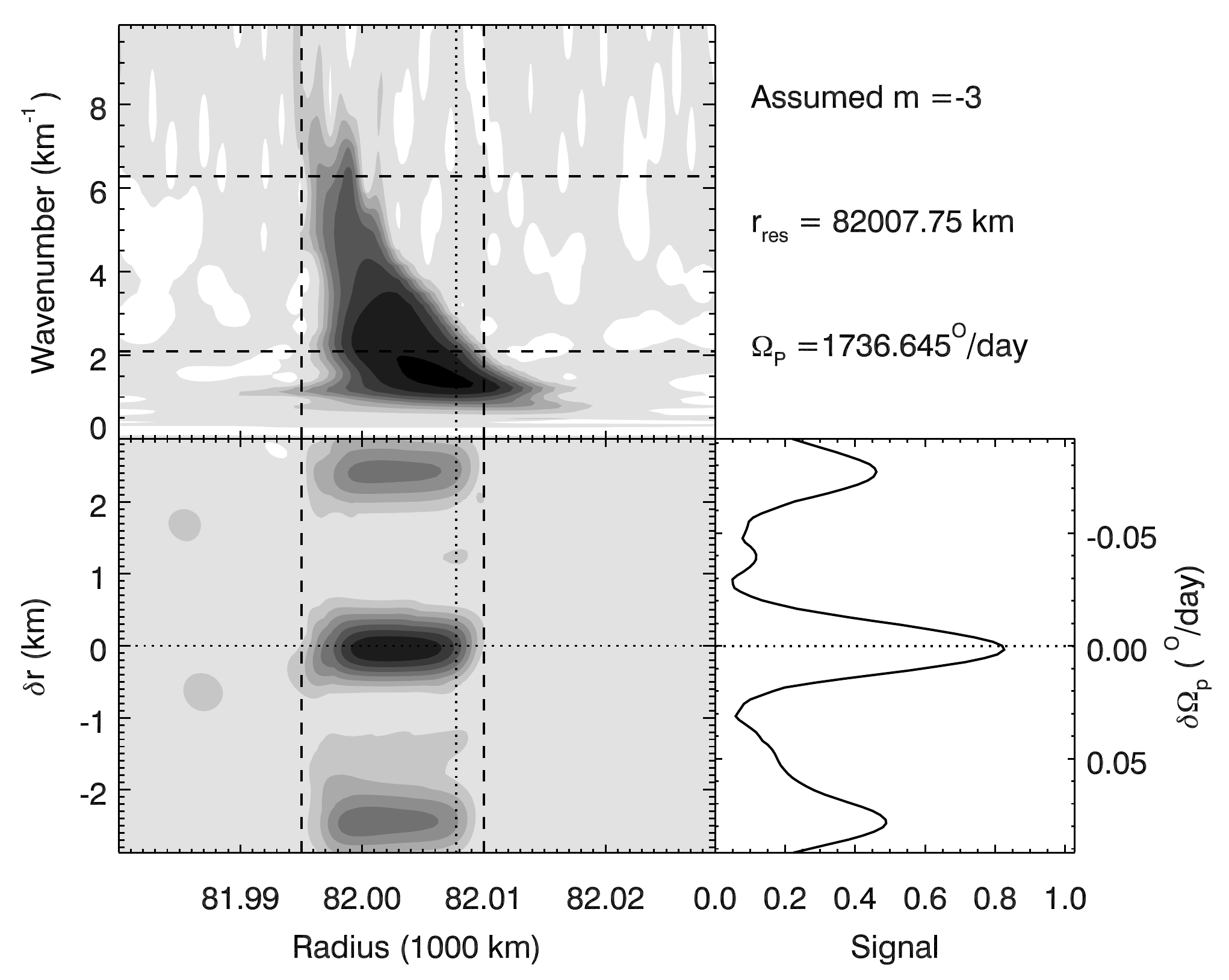}}
\caption{Detailed wavelet analysis of the $m=-3$ wave W82.00. See Figure~\ref{m2wave} for details. Note that the assumed pattern speed corresponds to the peak signal, and the wavelet power ratio shown in the top panel looks like a sensible inwardly-propagating density wave}
\label{m3awave}
\end{figure}

\begin{figure}
\resizebox{3.in}{!}
{\includegraphics{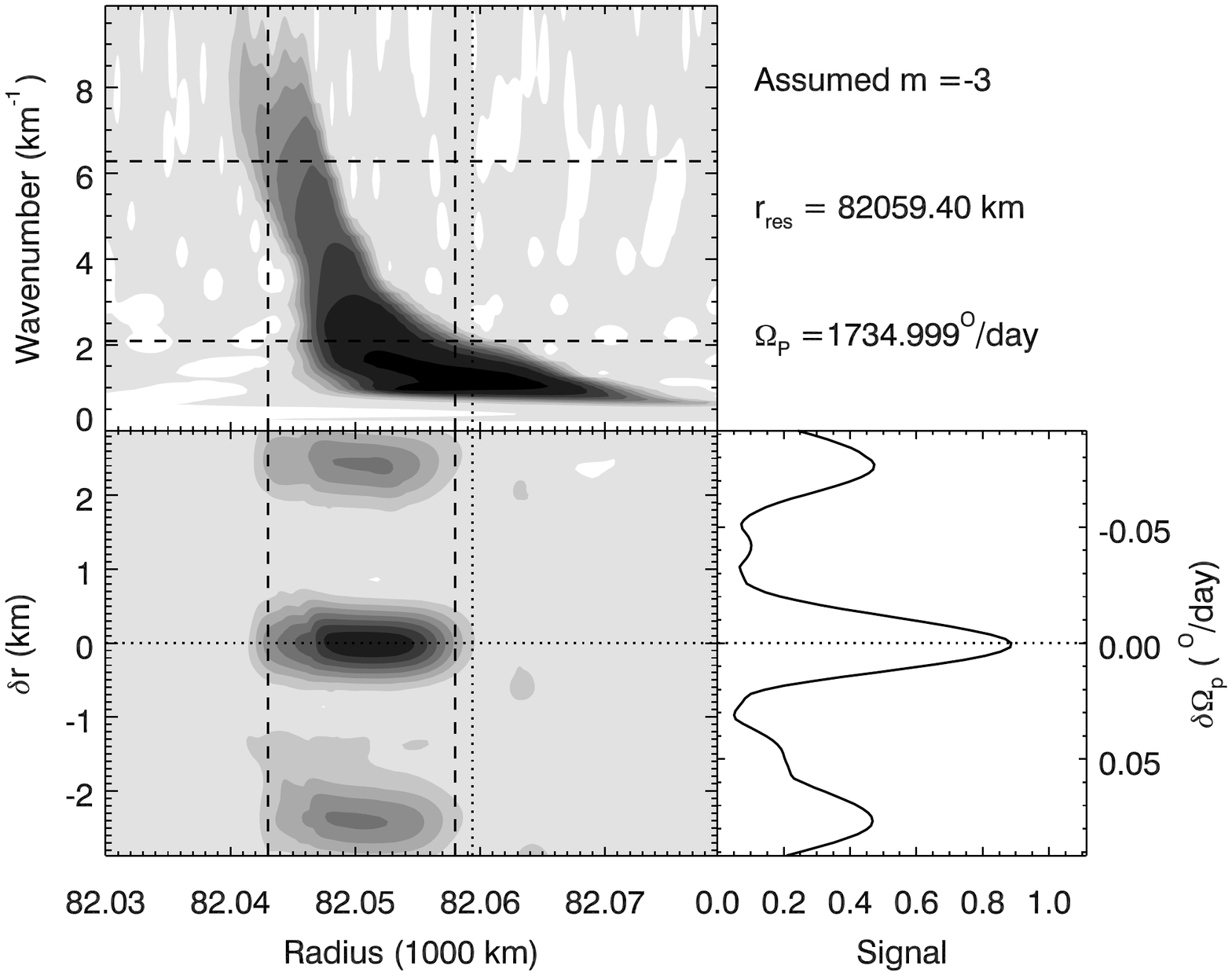}}
\caption{Detailed wavelet analysis of the $m=-3$ wave W82.06. See Figure~\ref{m2wave} for details. Note that the assumed pattern speed corresponds to the peak signal, and the wavelet power ratio shown in the top panel looks like a sensible inwardly-propagating density wave}
\label{m3bwave}
\end{figure}

\begin{figure}
\resizebox{3.in}{!}
{\includegraphics{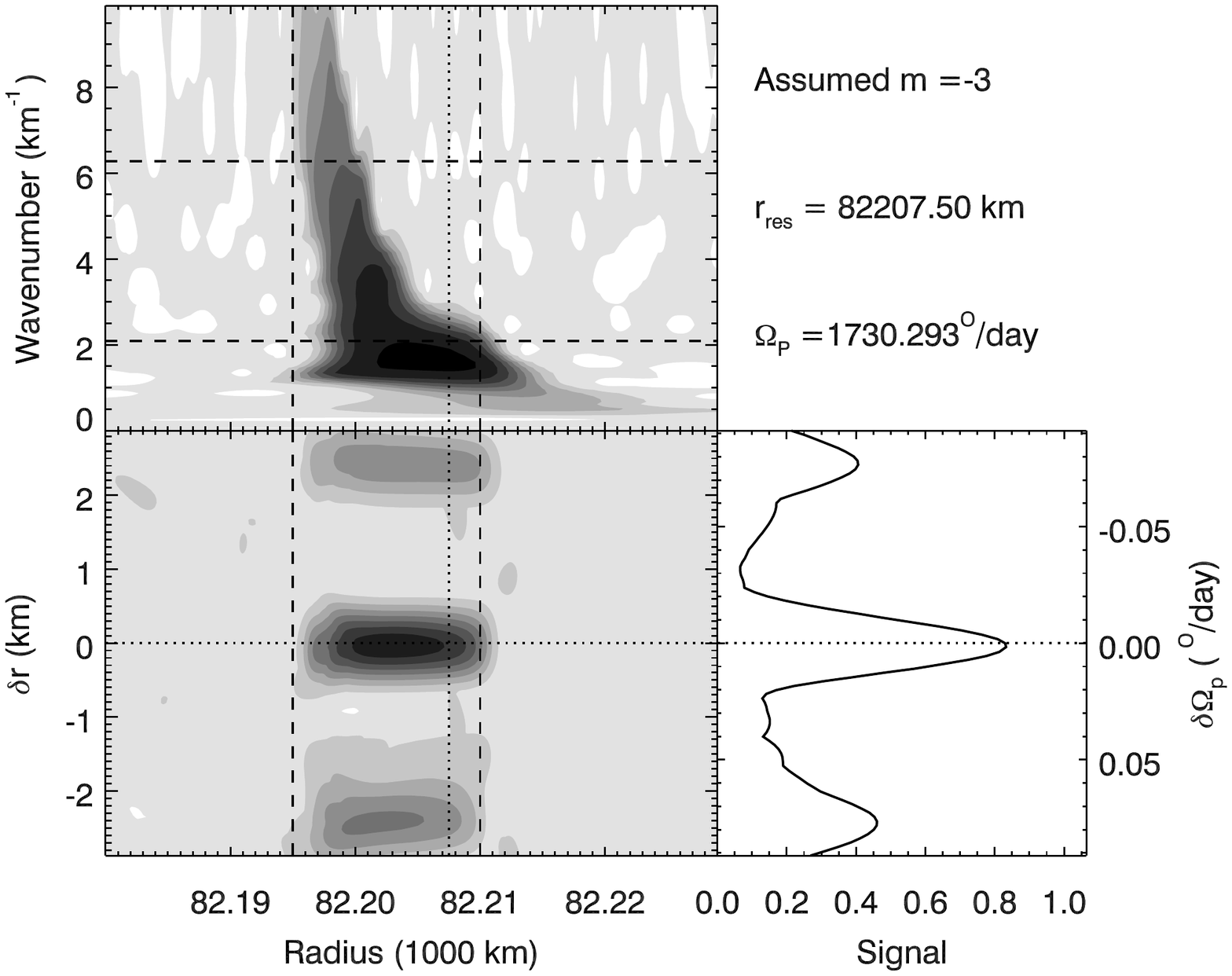}}
\caption{Detailed wavelet analysis of the $m=-3$ wave W82.21. See Figure~\ref{m2wave} for details. Note that the assumed pattern speed corresponds to the peak signal, and the wavelet power ratio shown in the top panel looks like a sensible inwardly-propagating density wave}
\label{m3cwave}
\end{figure}

\begin{figure}
\resizebox{3.in}{!}
{\includegraphics{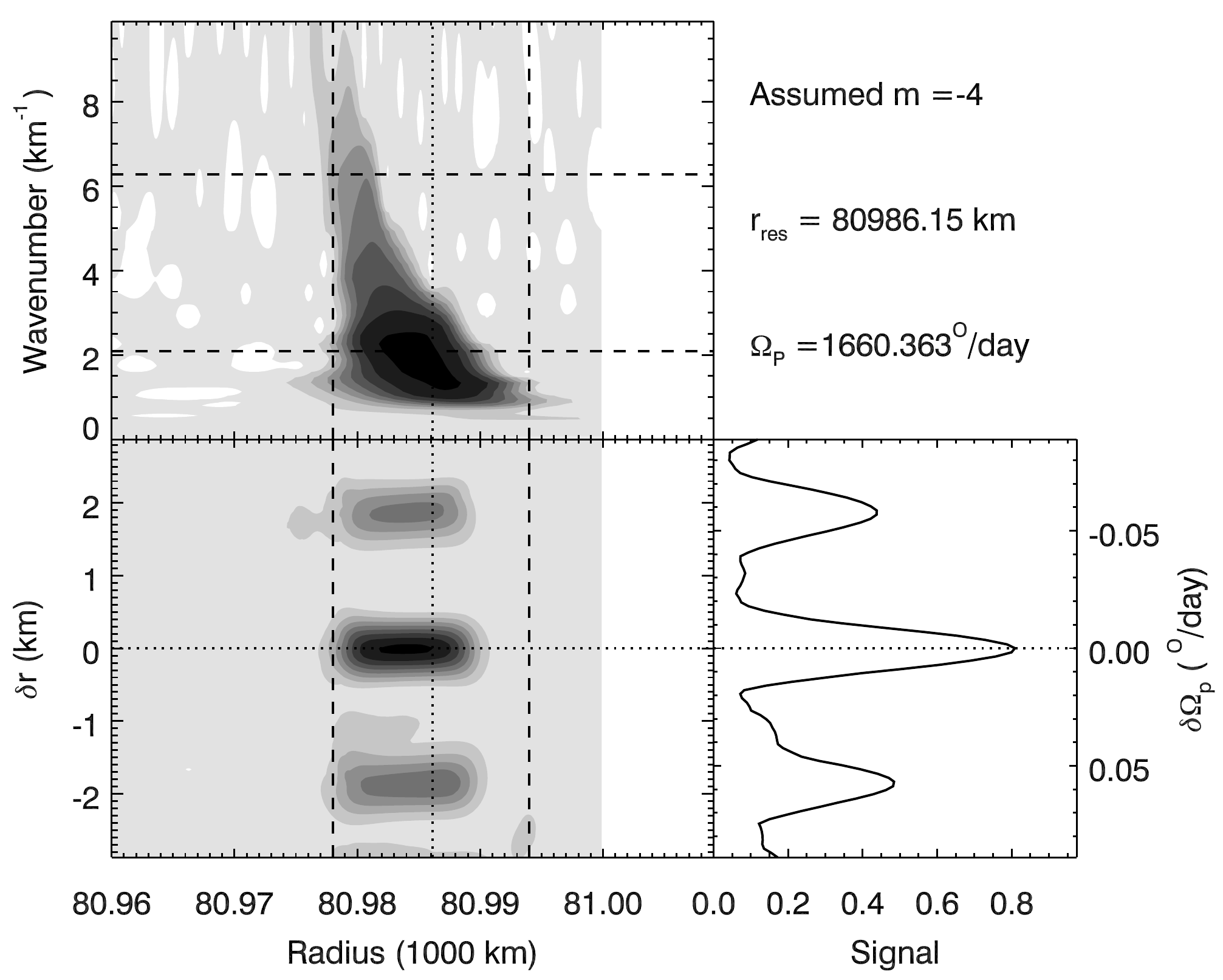}}
\caption{Detailed wavelet analysis of the $m=-4$ wave W80.98. See Figure~\ref{m2wave} for details. Note that the assumed pattern speed corresponds to the peak signal, and the wavelet power ratio shown in the top panel looks like a sensible inwardly-propagating density wave}
\label{m4wave}
\end{figure}

\begin{figure}
\resizebox{3.in}{!}
{\includegraphics{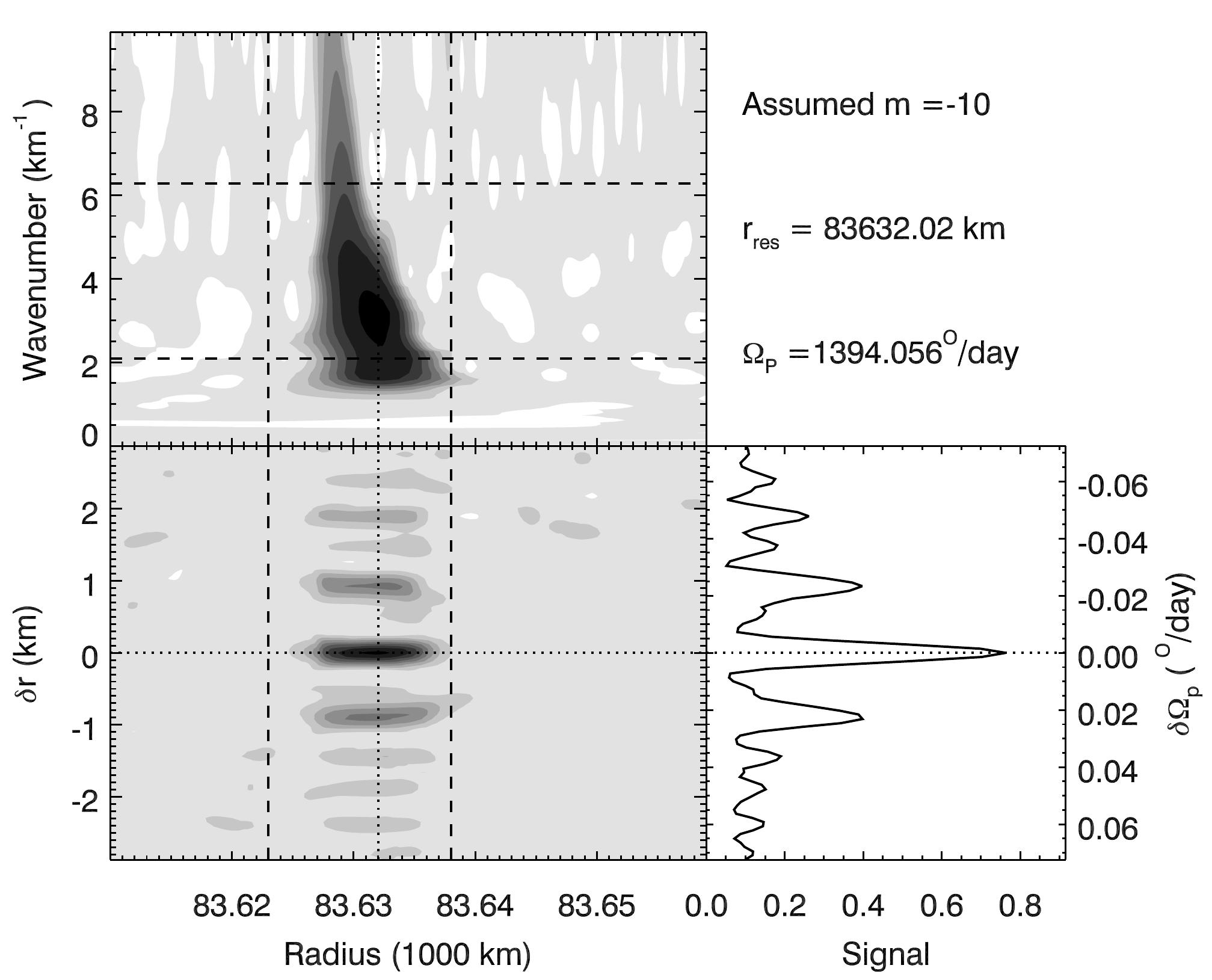}}
\caption{Detailed wavelet analysis of the $m=-10$ wave W83.63. See Figure~\ref{m2wave} for details. Note that the assumed pattern speed corresponds to the peak signal, and the wavelet power ratio shown in the top panel looks like a sensible inwardly-propagating density wave}
\label{m10wave}
\end{figure}

\begin{figure}
\resizebox{3.in}{!}
{\includegraphics{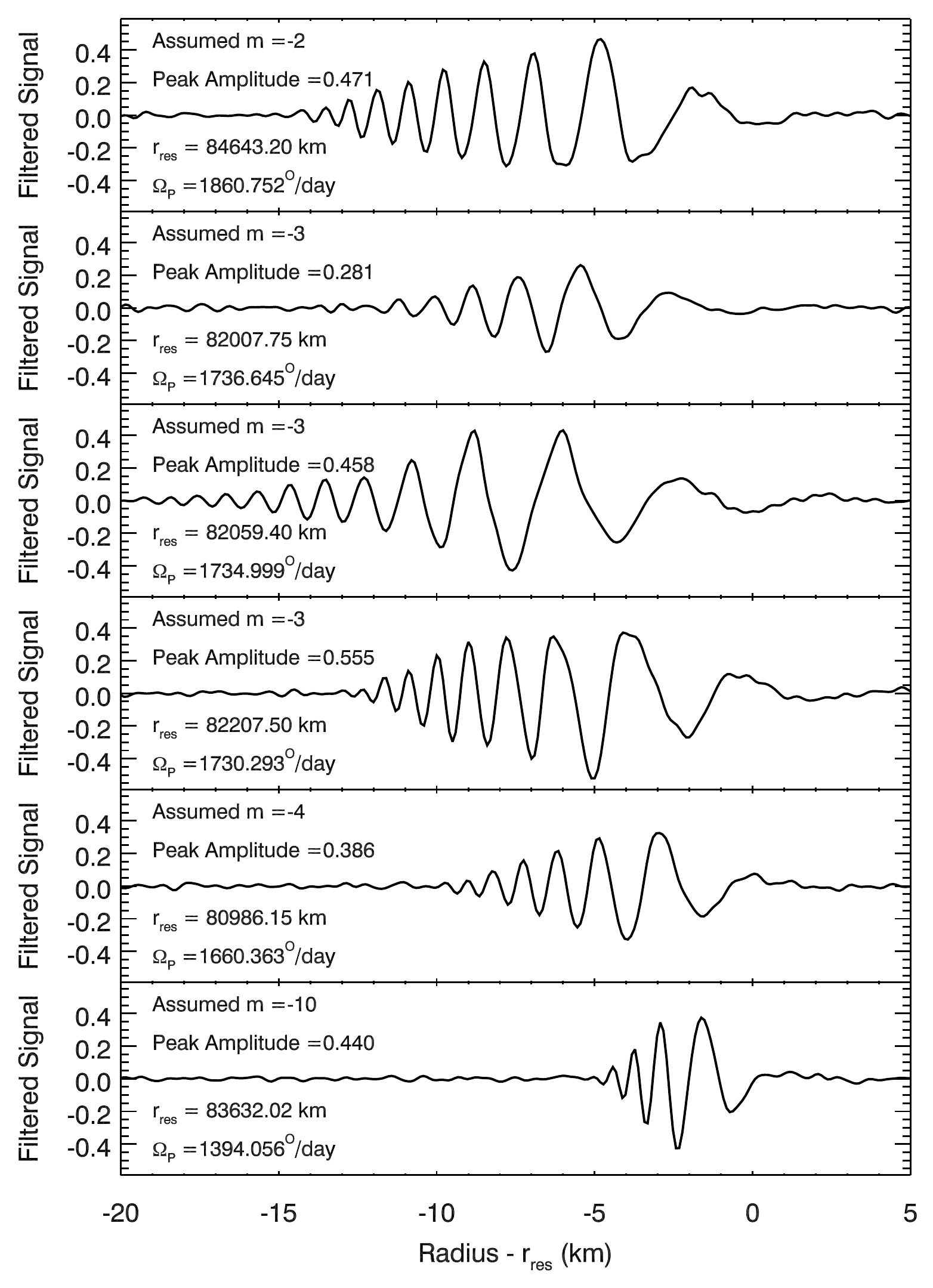}}
\caption{Reconstructed wave profiles for the waves identified in \citet{HN13,HN14} based on the full suite of occultations. These profiles are generated using the real part of the average phase-corrected wavelet between 0.5 km and 5 km. Note that the fractional optional depth variations associated with these waves are greater than 0.25, consistent with these waves being visible in individual profiles, and allowing many peaks to be observed in each profile.}
\label{recwave1}
\end{figure}

Figures~\ref{m2wave}-\ref{m10wave} show the results of the wavelet analysis for the previously identified waves W84.64, W82.00, W82.06, W82.21, W80.98 and W83.63. For the sake of conciseness, these figures just show the power ratio $\mathcal{R}$, the parameter that shows the wave signatures most clearly. The upper panels in these figures show $\mathcal{R}$ as a function of radius and wavenumber for the selected best-fit pattern speed, while the lower left-hand panels show the peak power ratio between wavelengths of 1 and 3 km as a function of radius and pattern speed (expressed as offsets in the assumed resonant radius {and pattern speed}). The lower right-hand panels show a profile of the maximum power ratio between wavelengths of 1 and 3 km  and between the two radii indicated by the vertical dashed lines. These profiles show clear peaks at the selected pattern speed for each wave, thus demonstrating that those pattern speeds best organize the signals associated with each wave. Many of these plots also show secondary maxima offset from the main peak by around 0.05$^\circ$/day. These peaks arise because the occultations are not evenly distributed in time, but instead come from three distinct time periods (2007-2010, 2012-2014 and 2016-2017) separated by 2-4 years. The secondary maxima correspond to one extra cycle of the pattern between these times. Fortunately, the three time periods are long enough and the gaps between them are short enough that there is no ambiguity in the best-fit solution for any of these waves. 

The width of the peaks also provides an estimate of the uncertainty in each wave's pattern speed. To be conservative, we will here report uncertainties that correspond to the full width of the peaks in the profiles. Hence W84.64 has a pattern speed of $1860.75\pm0.03^\circ$/day, W82.00 has a pattern speed of $1736.65\pm0.02^\circ$/day, W82.06 has a pattern speed of $1735.00\pm0.02^\circ$/day, W82.21 has a pattern speed of $1730.30\pm0.02^\circ$/day, W80.98 has a pattern speed of $1660.36\pm0.02^\circ$/day, and W83.63  has a pattern speed of $1394.06\pm0.01^\circ$/day. It is important to note that the actual peak positions can be determined to significantly higher precision than this. However, providing statistically rigorous uncertainties on these parameters is not practical at this time because we are considering maximum values of the power ratio over ranges of radii and wavenumbers, and propagating the appropriate uncertainties is beyond the scope of this report.

Figure~\ref{recwave1} shows the reconstructed wave profiles derived from the average phase-corrected wavelets, specifically wavelengths between 0.5  and 5 km. Note that these profiles have been normalized so that they represent the fractional optical depth variations associated with each wave.  The peak amplitudes of all these waves are all greater than 0.25, so the optical depth variations associated with these waves are a large fraction of the mean optical depth. Standard linear density wave theory is therefore not strictly appropriate for these waves. The nonlinear aspects of these waves do not affect their symmetry properties and pattern speeds, but do affect the  detailed shape of the density variations (for example, they give rise to the asymmetries in the shapes of individual peaks and troughs). More importantly, the standard expressions relating the wave amplitude to the strength of the applied perturbation do not hold, complicating any effort to translate the relative amplitudes of these waves into information about the relative amplitudes of the oscillations inside the planet.

\begin{figure}
\resizebox{3.in}{!}
{\includegraphics{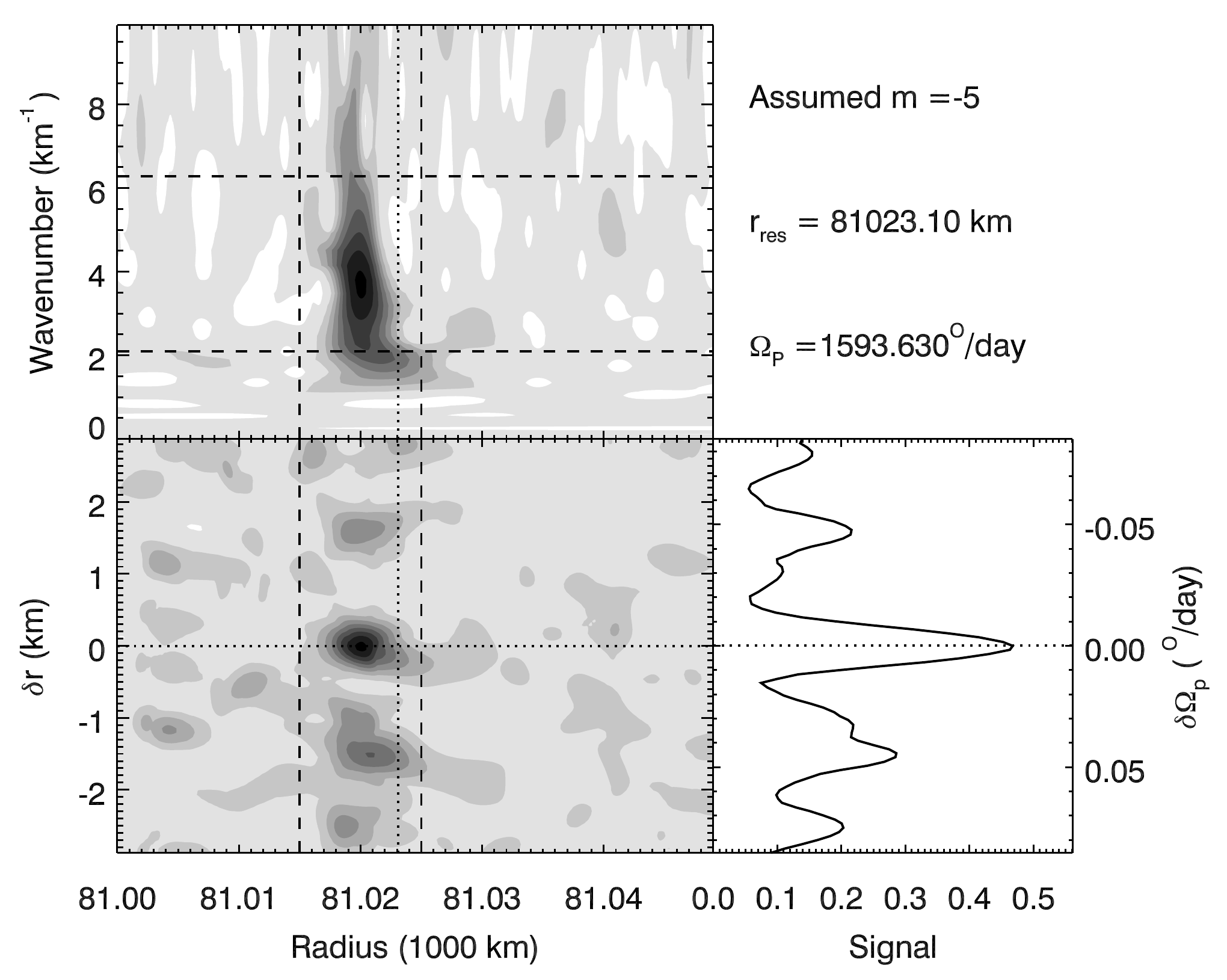}}
\caption{Detailed wavelet analysis of the $m=-5$ wave W81.02a. See Figure~\ref{m2wave} for details. Note that the assumed pattern speed corresponds to the peak signal, and the wavelet power ratio shown in the top panel looks like a sensible inwardly-propagating density wave}
\label{m5wave}
\end{figure}

\begin{figure}
\resizebox{3.in}{!}
{\includegraphics{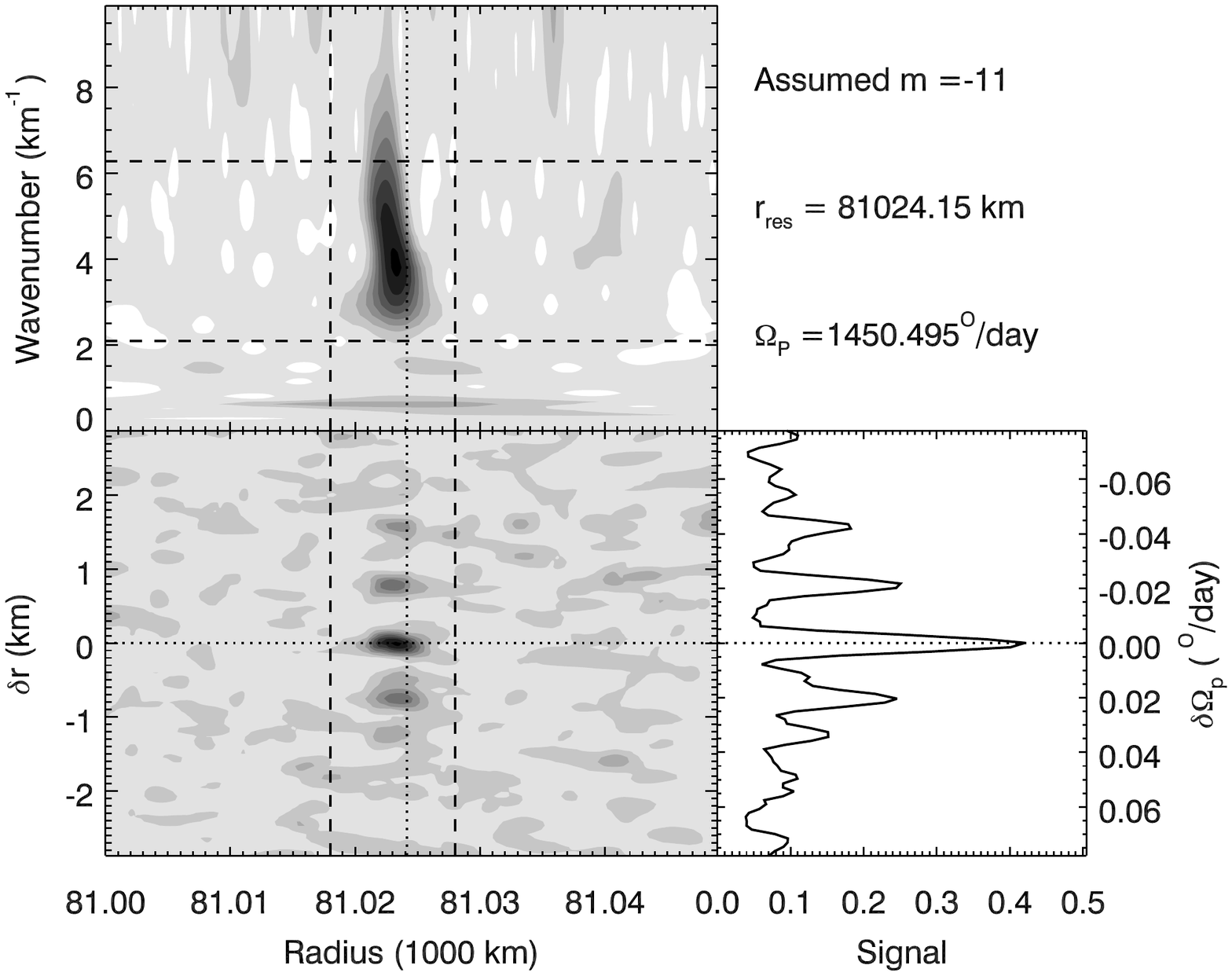}}
\caption{Detailed wavelet analysis of the $m=-11$ wave W81.02b. See Figure~\ref{m2wave} for details. Note that the assumed pattern speed corresponds to the peak signal, and the wavelet power ratio shown in the top panel looks like a sensible inwardly-propagating density wave}
\label{m11wave}
\end{figure}

\bigskip

\bigskip

\subsection{The $m=-5$ and $m=-11$ components of W81.02}

The results of the wavelet analyses of the two components of the W81.02 wave are shown in Figures~\ref{m5wave} and~\ref{m11wave}. Consistent with the analysis presented in Section~\ref{W81p02}, there are clear signals in the power ratios for both $m=-5$ and $m=-11$. The $m=-5$ component has a peak signal at a pattern speed of 1596.63$\pm0.02^\circ/$day, while the $m=-11$ component has a peak signal at a pattern speed of $1450.50\pm0.01^\circ/$day. (Again, the uncertainties are conservative estimates  based on the full widths of the peaks in the power ratio profiles). These two pattern speeds correspond to resonant radii separated by only about 1 km. Also, the signals seen at the appropriate pattern speeds have sensible trends in wavenumber-radius space, with the strongest signals being seen just interior to the nominal resonance location, and the signal occurring at higher wavenumbers further inwards from the resonance.

\begin{figure}
\resizebox{3.in}{!}
{\includegraphics{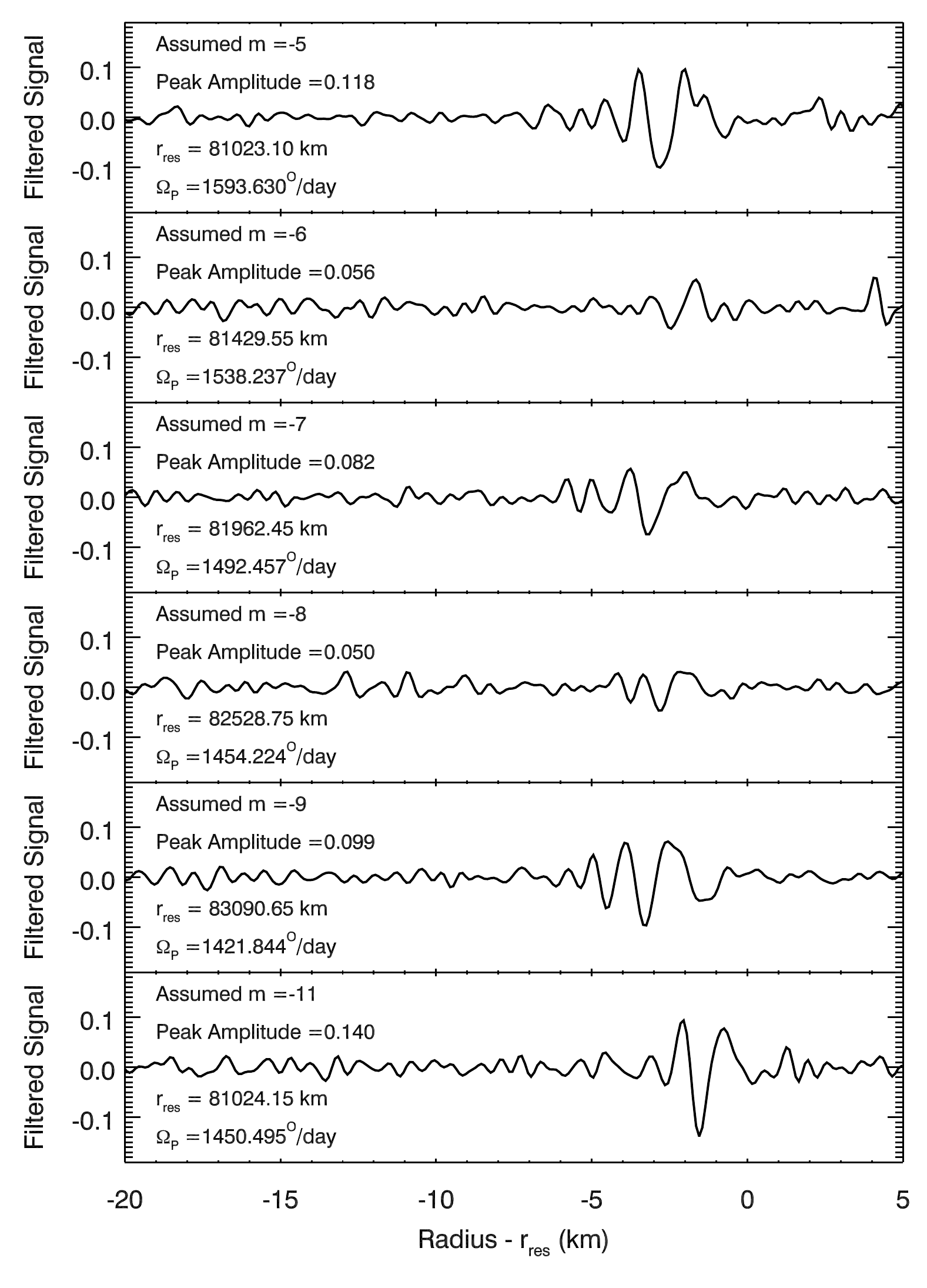}}
\caption{Reconstructed wave profiles for the newly identified waves based on the full suite of occultations. These profiles are generated using the real part of the average phase-corrected wavelets between 0.5 km and 5 km.}
\label{recwave2}
\end{figure}

The reconstructed wave profiles for both these waves derived from  the Group C occultations are shown in the top and bottom panels of Figure~\ref{recwave2}. These waves clearly have much lower signal-to-noise than any of the waves described in the previous subsection, and only a few wave cycles are visible for each of these waves. Still, as will be demonstrated below, these signals are consistently found among sub-sets of the data with peak amplitudes between 0.1 and 0.15. Hence W81.02a and W81.02b are both valid waves, but are also weaker than any of the previously identified waves.

\subsection{The $m=-7$ and $m=-9$ wave signatures W81.96 and W83.09}

\begin{figure}
\resizebox{3.in}{!}
{\includegraphics{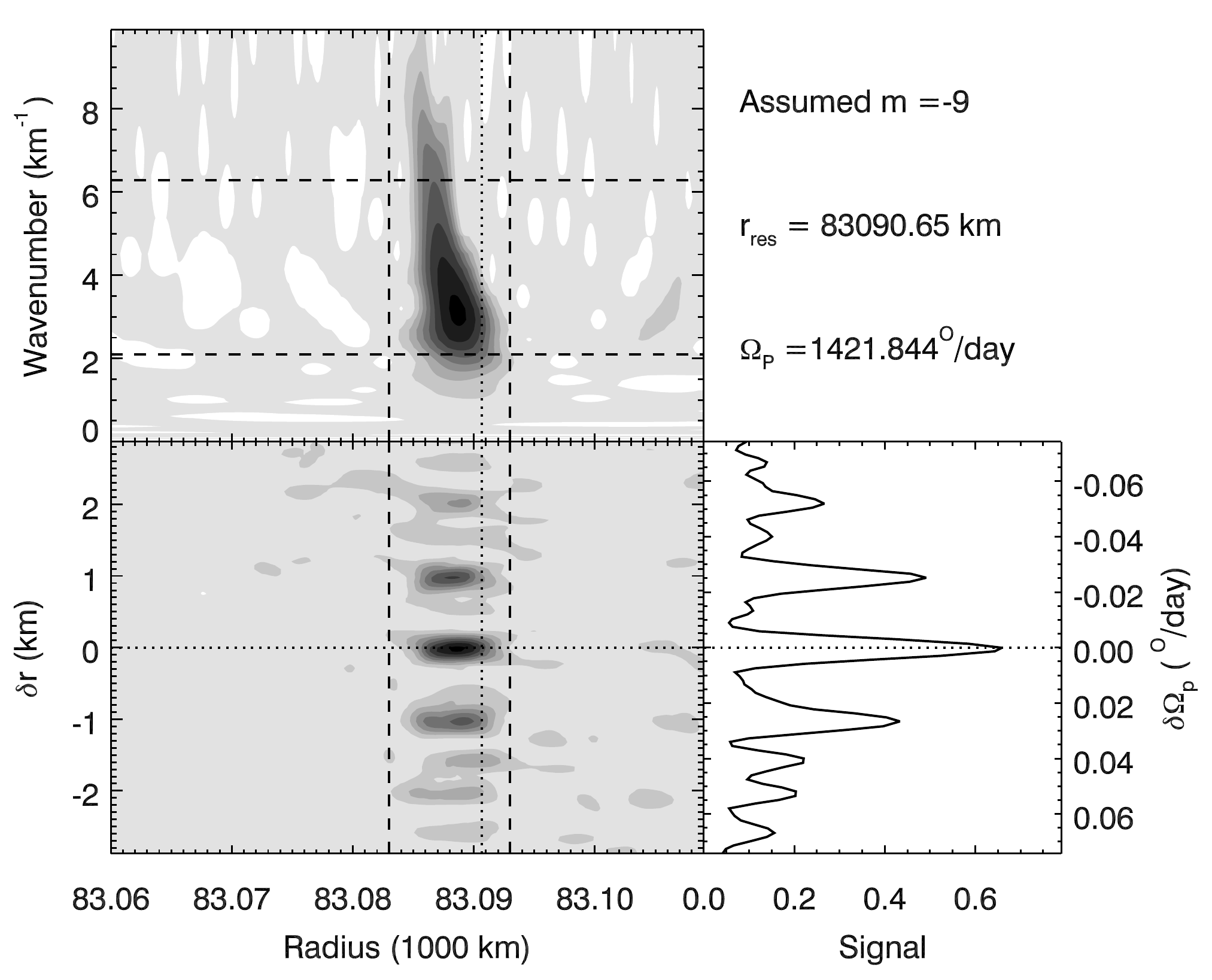}}
\caption{Detailed wavelet analysis of the $m=-9$ wave W83.09. See Figure~\ref{m2wave} for details. Note that the assumed pattern speed corresponds to the peak signal, and the wavelet power ratio shown in the top panel looks like a sensible inwardly-propagating density wave}
\label{m9wave}
\end{figure}

Among the four new wave signatures revealed by our search, the $m=-7$ and $m=-9$ signals designated W81.96 and W83.09 are stronger and more robust, and so we will consider them first.  Figure~\ref{m9wave} shows the wavelet analysis of the $m=-9$ signal W83.09, which shows a clear peak at 1421.84$\pm0.01^\circ$/day, corresponding to a resonant radius of 83090.65 km. The signal is also perfectly consistent with an inward-propagating density waves, falling just inside the expected resonant radius and showing a clear increase in wavenumber with distance from the resonance. Furthermore, the reconstructed wave profile for this signal, shown in Figure~\ref{recwave2}, looks like a sensible inward-propagating density wave with a peak amplitude of around 0.10, which is not much smaller than the components of W81.02. We can therefore be fairly confident that this signal comes from a real  $m=-9$ density wave.

\begin{figure}
\resizebox{3.in}{!}
{\includegraphics{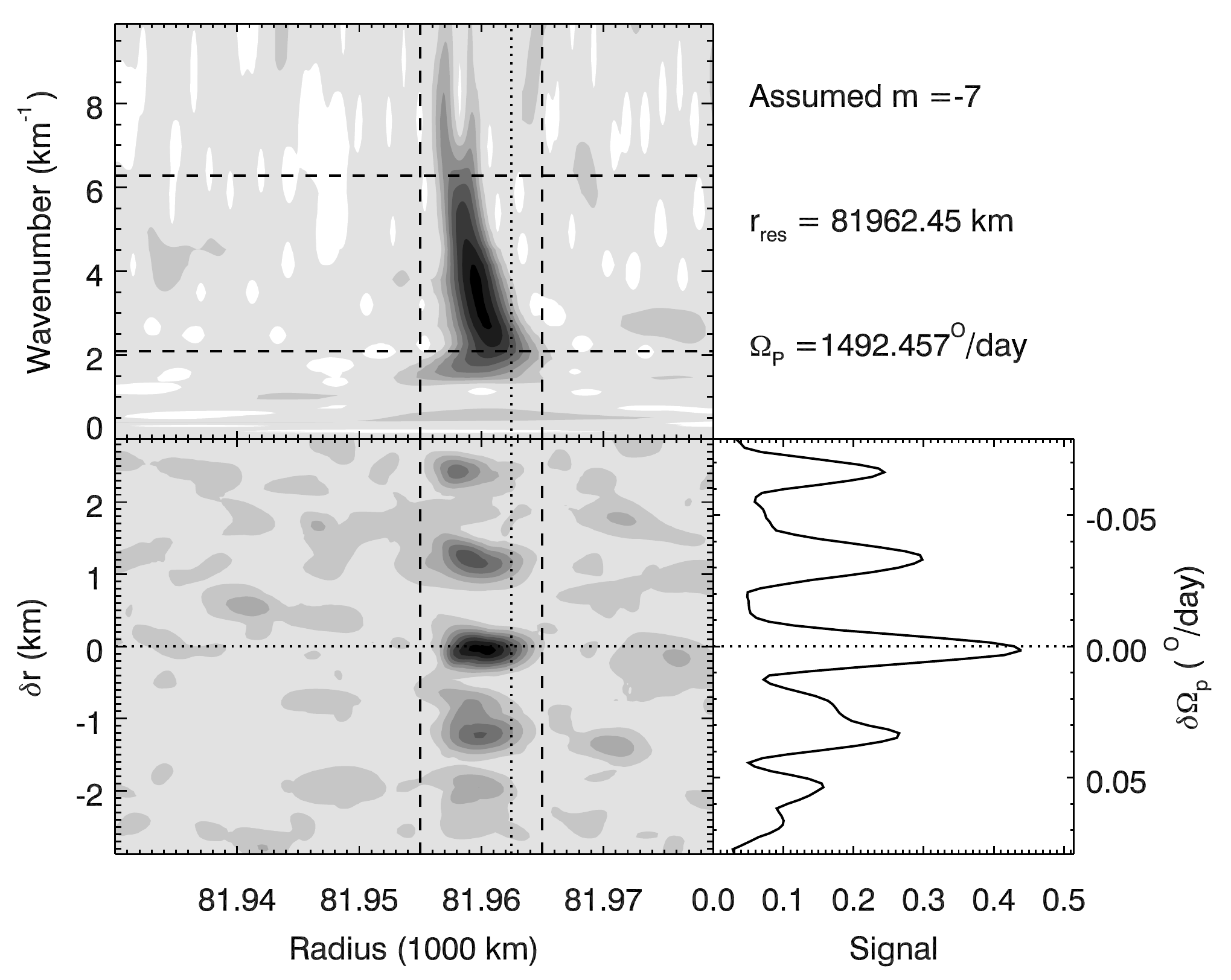}}
\caption{Detailed wavelet analysis of the $m=-7$ wave W81.96. See Figure~\ref{m2wave} for details. Note that the assumed pattern speed corresponds to the peak signal, and the wavelet power ratio shown in the top panel looks like a sensible inwardly-propagating density wave}
\label{m7wave}
\end{figure}

The wavelet analysis of W81.96 shows a comparably clear  $m=-7$ signal with a pattern speed of 1492.46$\pm0.02^\circ/$day, corresponding to a resonant radius of 81962.45 km (see Figure~\ref{m7wave}). This signal also appears to be quite consistent with an inwardly-propagating density wave, with another clear trend where the wavenumber increases inwards. The reconstructed wave profile for W81.96, while having a slightly lower amplitude than W83.09, still preserves multiple wave cycles and looks like a reasonable  inward-propagating density wave. Thus we can conclude that W81.96 is indeed an $m=-7$ density wave.

\subsection{The $m=-6$ and $m=-8$ wave candidates W81.43 and W82.53}

\begin{figure}
\resizebox{3.in}{!}
{\includegraphics{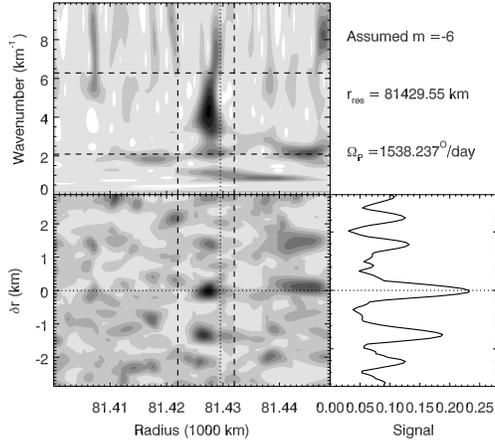}}
\caption{Detailed wavelet analysis of the $m=-6$ wave W81.43. See Figure~\ref{m2wave} for details. Note that the assumed pattern speed corresponds to the peak signal, but the wavelet power ratio is not clearly a sensible inwardly-propagating density wave}
\label{m6wave}
\end{figure}

\begin{figure}
\resizebox{3.in}{!}
{\includegraphics{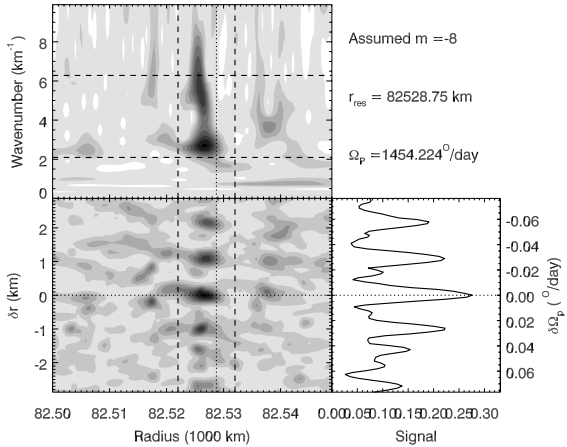}}
\caption{Detailed wavelet analysis of the $m=-8$ wave W82.53. See Figure~\ref{m2wave} for details. Note that the assumed pattern speed corresponds to the peak signal, but there are multiple other pattern speeds with comparable signal levels.}
\label{m8wave}
\end{figure}

\begin{figure}
\resizebox{3.in}{!}{\includegraphics{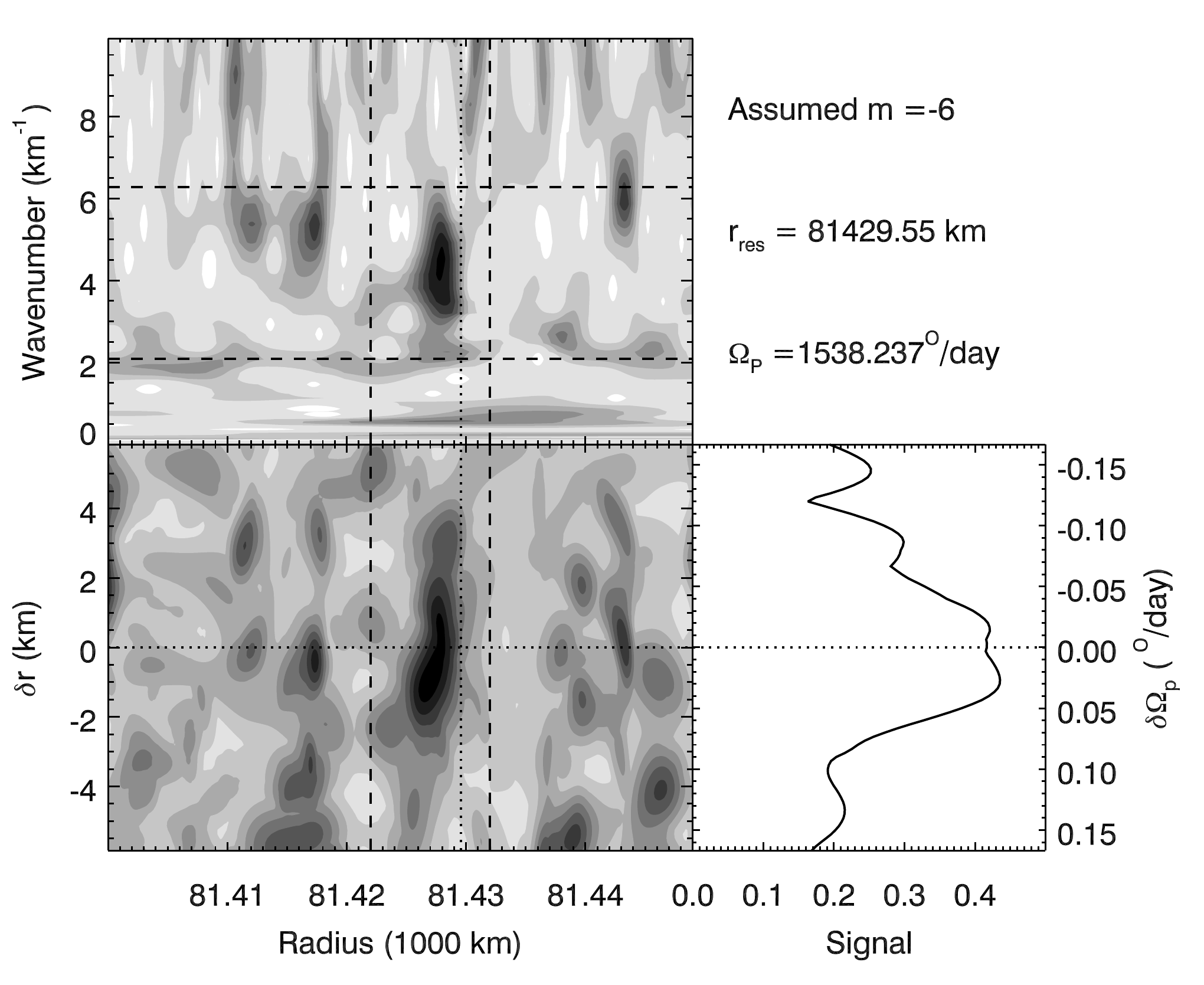}}
\resizebox{3.in}{!}
{\includegraphics{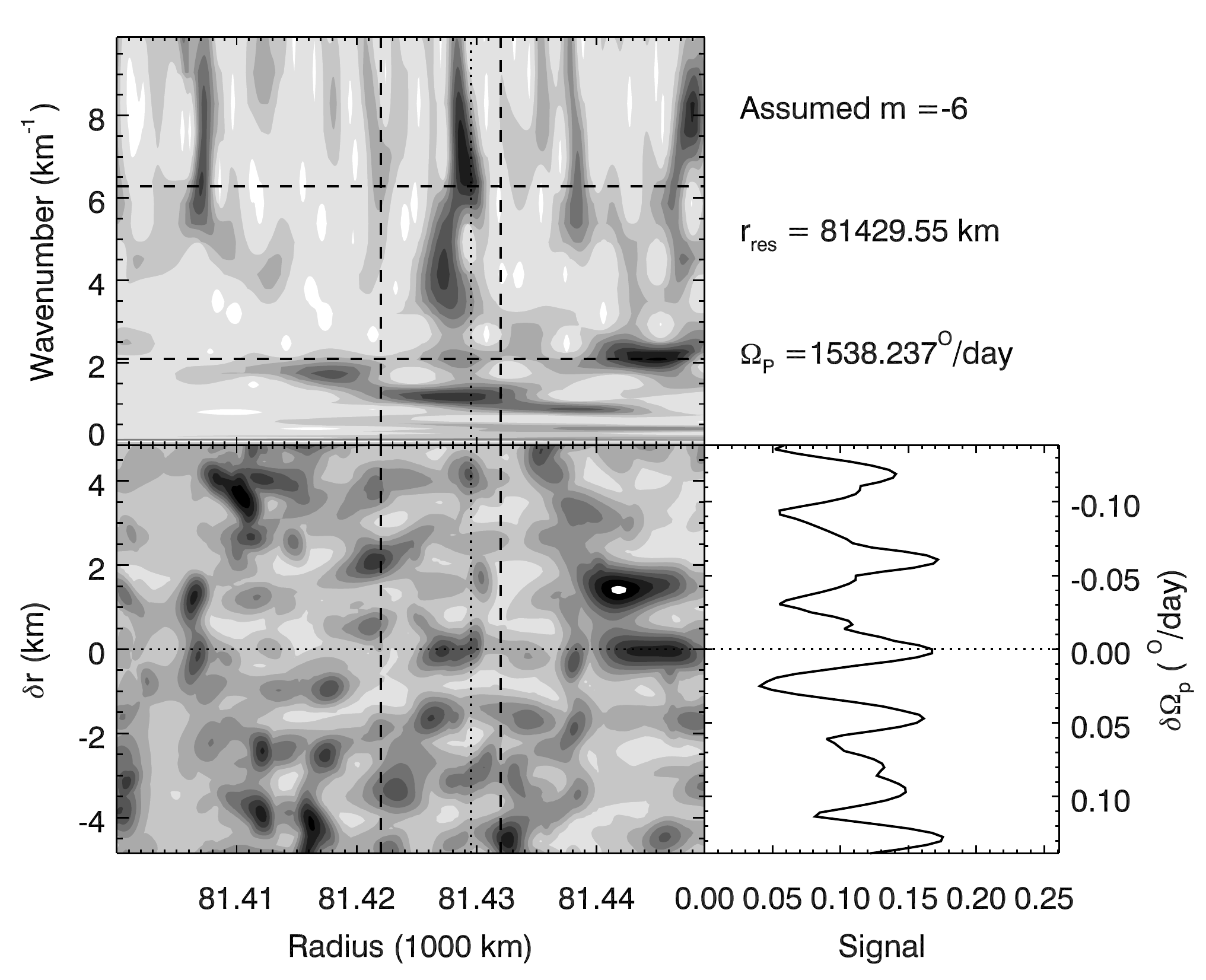}}
\caption{Detailed wavelet analysis of the $m=-6$ wave W81.43, using only data obtained before 2010 (top) or after 2010 (below). See Figure~\ref{m2wave} for details.  Note that both analyses use the same reference pattern speed that best-fit the full data set (Figure~\ref{m6wave}), and both power ratios show a signal at the same combination of radii, pattern speeds and wavenumbers.}
\label{m6wave2}
\end{figure}

\begin{figure}
\resizebox{3.in}{!}{\includegraphics{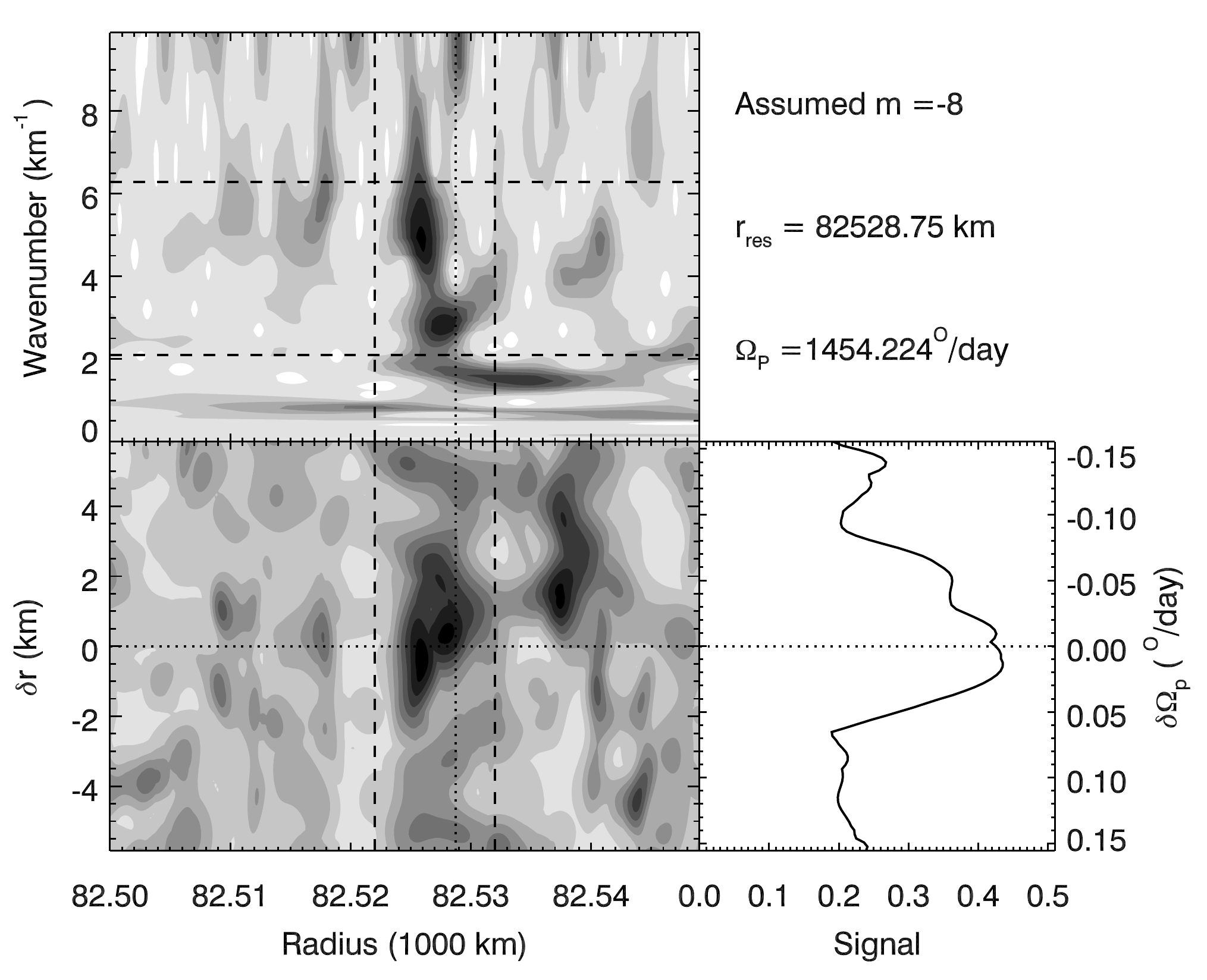}}
\resizebox{3.in}{!}
{\includegraphics{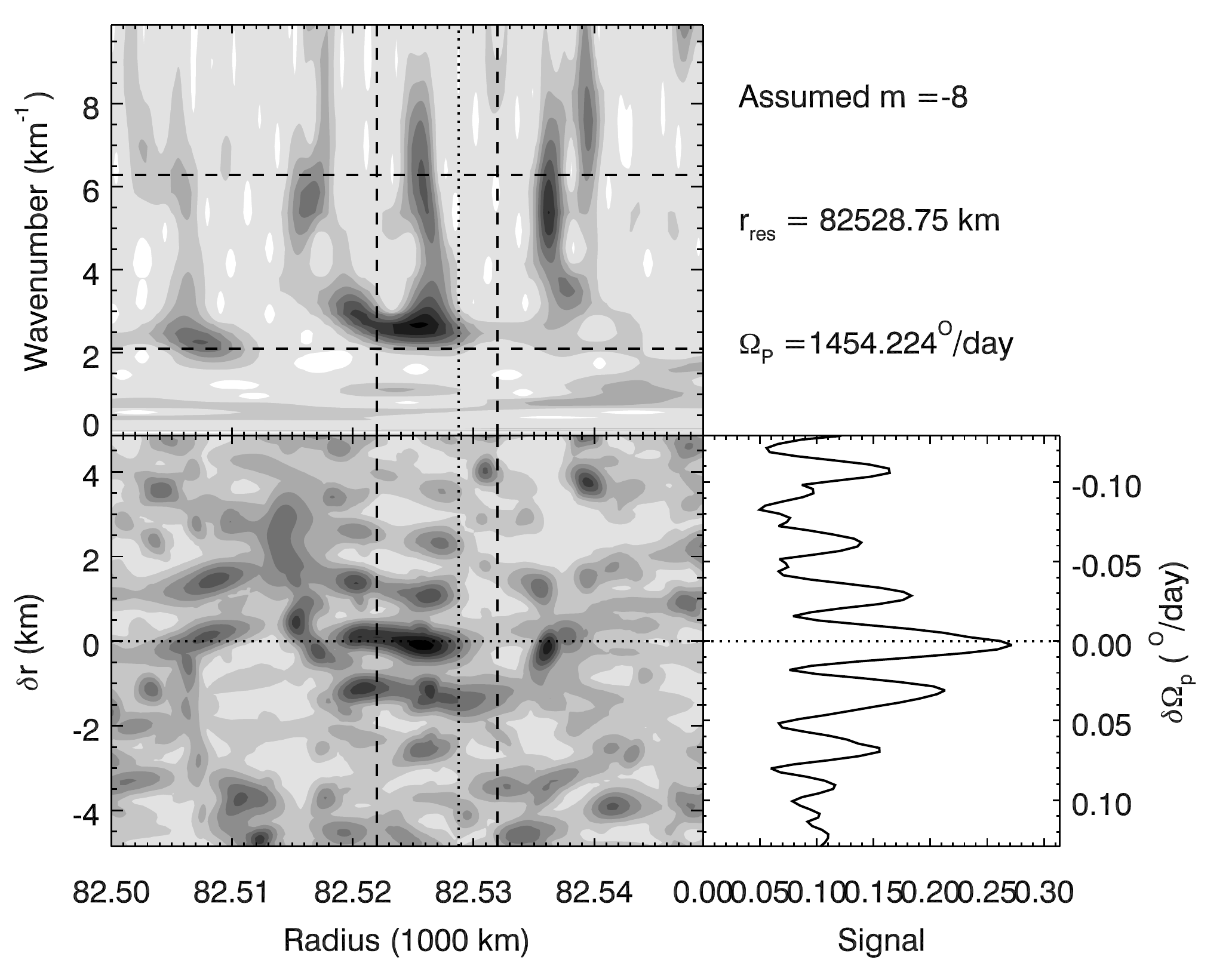}}
\caption{Detailed wavelet analysis of the $m=-8$ wave W82.53, using only data obtained before 2010 (top) or after 2010 (below). See Figure~\ref{m2wave} for details.  Note that both analyses use the same reference pattern speed that best-fit the full data set (Figure~\ref{m8wave}), and both power ratios show a signal at the same combination of radii, pattern speeds and wavenumbers.}

\label{m8wave2}
\end{figure}

\begin{figure}
\resizebox{3.in}{!}{\includegraphics{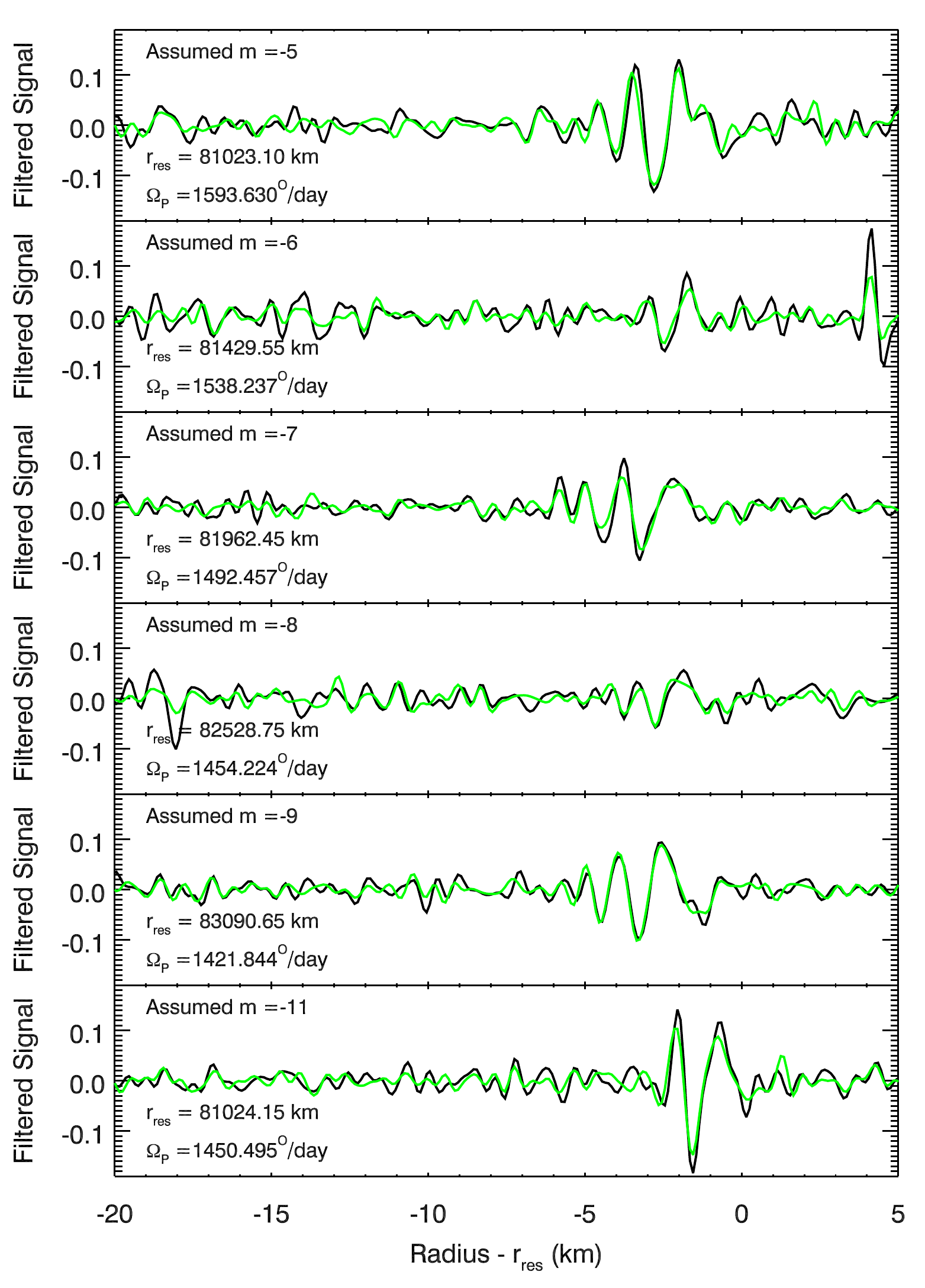}}
\caption{Reconstructed wave profiles for the waves identified in this paper based occultations obtained before  and after 2010 in black and green, respectively. These profiles are generated using the part of the average phase-corrected wavelet between 0.5 km and 5 km. In all cases, the variations associated with the wave candidate show similar wavelengths and phases.}
\label{recwave2a}
\end{figure}

Finally, we must consider the candidate $m=-6$ and $m=-8$ waves W81.43 and W82.53. Figures~\ref{m6wave} and~\ref{m8wave} show wavelet analyses of these signals based of the full suite of occultations. For W81.43 there does appear to be a peak is the power ratio when $m=-6$ at a pattern speed of 1538.24$\pm0.02^\circ$/day, corresponding to a resonant radius of 81429.55 km, which falls just outside the region with the strongest signal. For W82.53 there are multiple peaks in the power ratio profile, but the strongest falls at 1454.22$\pm0.01^\circ$/day, which corresponds to a resonant radius of 82528.75 km, a sensible location just outside the observed signal. The signal associated with this pattern speed also exhibits sensible trends in wavenumber-radius space, with higher wavenumbers being found at increasing distances from the resonance.

Turning to the reconstructed wave profiles for these regions (shown in Figure~\ref{recwave2}) we find that both these waves have extremely small amplitudes, and are just barely above the background noise fluctuations. W82.53 appears to be a scaled down version of the other waves, and so is perhaps more convincing. By contrast, W81.43 only preserves a cycle or two, and so does not look particularly wavelike. In both cases, one can reasonably ask whether these are real wave signatures or just a chance alignment of random noise in the various profiles. 

To address these concerns, we sought to establish whether these two signals could be seen throughout the Cassini mission. We therefore divided the occultations into two groups based on whether they were observed before or after 2010. This produced two separate data sets with roughly comparable signal-to-noise. Figures~\ref{m6wave2} and~\ref{m8wave2} show the results of the wavelet analyses for these two time periods for each of the wave candidates W81.43 and W82.53, while Figure~\ref{recwave2a} shows the reconstructed profiles derived for all the weak waves derived from these two time periods.

For both W81.43 and W82.53, the data obtained before 2010 shows a relatively clear peak in the power ratio at the expected pattern speed that strongly resembles the signal seen in the full data set. By contrast, the data obtained after 2010 do not show such a unique signal. While the expected pattern speed does correspond to a peak in the power ratios, there are multiple peaks of comparable strength at other pattern speeds and locations. The strongest signals therefore appear to be restricted to the early part of the Cassini mission. However, both the early and late data for each wave candidate do show signals at the same radii and wavenumbers for the selected pattern speeds. This at least hints that the signal seen prior to 2010 did persist to later times. Also, if one examines the reconstructed profiles for these two time periods (shown in Figure~\ref{recwave2a}), one finds that the individual peaks and troughs associated with these wave candidates do line up, unlike the other features in these profiles. Furthermore, while the amplitude of the structures seen after 2010 are lower than those seen before 2010, this appears to be a general trend common to all the wave signals.  

 {It is still unclear why the wave signals seem to be stronger in the earlier data. As shown in Table~\ref{obstab}, many of the occultations observed before 2010 used the star $\gamma$ Crucis. While the high elevation angle of $\gamma$ Crucis above the rings makes these occultations especially useful for probing high-optical regions like the B ring \citep{HN16}, it is not so obvious what would make $\gamma$ Crucis occultations especially sensitive to waves in low-optical depth regions like the C ring. It could be that the more heterogeneous occultations used later in the mission somehow reduced the efficacy of the phase-correction techniques, but we have thus far been unable to identify any evidence for this. Hence, the question of whether these differences reflect a subtle artifact of our data processing algorithms or a real temporal variation in the waves, must be left as an open question for future work.}

The evidence for W81.43 and W82.53 being real density waves is certainly weaker than the other signals considered in this report. However, the relatively unambiguous signals in the pre-2010 data, along with the consistent patterns seen before and after 2010 are sufficient for us to regard these features as reasonable candidate $m=-6$ and $m=-8$ density waves.


\begin{table*}
\caption{Summary of wave properties}
\label{wavetab}
\hspace{-1in}\resizebox{7.5in}{!}{\begin{tabular}{|c|c|c|c|c|c|c|c|c|c|}\hline
Wave Name & Other & Previous & Figures & Radii Considered & Wavelengths  & $m$ & 
Resonant  & Pattern Speed & Peak Wave \\
&  Designations$^a$ & Analysis$^b$ &  & & Considered & &  Radius$^d$ & & Amplitude$^e$ \\ \hline
W80.98 & Bailli\'e 13, Rosen e & HN13 & \ref{m4wave}, \ref{recwave1}  & 80978-80994 km & 1-3 km & $-4$ &80986.15 km & 1660.36$\pm0.02^\circ$/day & 0.386 \\
W81.02a& Bailli\'e 14  & HN14$^c$ &  \ref{m5wavesep}, \ref{m5wave}, \ref{recwave2}, \ref{recwave2a}  & 81016-81025 km &1-3 km & $-5$ & 81023.15 km & 1593.63$\pm0.02^\circ$/day & 0.118 \\
W81.02b&Bailli\'e 14 & HN14$^c$ &  \ref{m11wavesep}, \ref{m11wave}, \ref{recwave2}, \ref{recwave2a} & 81019-81028 km &1-3 km & $-11$ & 81024.17 km & 1450.50$\pm0.01^\circ$/day & 0.140 \\
W81.43 & & &  \ref{recwave2}, \ref{m6wave}, \ref{m6wave2}, \ref{recwave2a} & 81422-81432 km &1-3 km & $-6$ & 81429.55 km & 1538.24$\pm0.04^\circ$/day & 0.056 \\
W81.96 & & &  \ref{recwave2}, \ref{m7wave}, \ref{recwave2a} & 81955-81965 km &1-3 km & $-7$ & 81962.45 km &  1492.46$\pm0.02^\circ$/day & 0.082 \\
W82.00 & Bailli\'e 15 & HN13, HN14 & \ref{m3awave}, \ref{recwave1} & 81995-82010 km &1-3 km & $-3$ & 82007.75 km &1736.65$\pm0.02^\circ$/day & 0.281 \\
W82.06 & Bailli\'e 16, Rosen f & HN13, HN14 & \ref{m3bwave}, \ref{recwave1} & 82043-82058 km &1-3 km & $-3$ & 82059.40 km &1735.00$\pm0.02^\circ$/day & 0.458 \\
W82.21 & Bailli\'e 17, Rosen g & HN13, HN14 & \ref{m3cwave}, \ref{recwave1} & 82195-82210 km &1-3 km & $-3$ & 82207.50 km &1730.29$\pm0.02^\circ$/day & 0.555 \\
W82.53 & & & \ref{recwave2}, \ref{m8wave}, \ref{m8wave2}, \ref{recwave2a} & 82522-82532 km &1-3 km & $-8$ & 82528.75 km &1454.22$\pm0.04^\circ$/day & 0.050 \\
W83.09 & & & \ref{recwave2}, \ref{m9wave} ,\ref{recwave2a} & 83038-83093 km &1-3 km & $-9$ & 83090.65 km &1421.84$\pm0.01^\circ$/day & 0.099 \\
W83.63 & Bailli\'e 18, Rosen h & HN14 & \ref{m10wave}, \ref{recwave1}  & 83623-83628 km &1-3 km & $-10$ &83632.02 km &1394.06$\pm0.01^\circ$/day & 0.440 \\
W84.64 & Bailli\'e 19, Rosen i & HN13, HN14 & \ref{m2wave}, \ref{recwave1}  & 84630-84640 km &1-3 km & $-2$ & 84643.20 km &1860.75$\pm0.03^\circ$/day & 0.471 \\
\hline
\end{tabular}}

$^a$ Designations from \citet{Baillie11} and \citet{Rosen91}

$^b$ HN13 = \citet{HN13}, HN14=\citet{HN14}

$^c$ Discussed, but not identified

$^d$ Derived from pattern speed

$^e$ No error is provided on these quantities because their uncertainties are dominated by systematic errors in the reconstruction that are difficult to rigorously quantify.
\end{table*}


 \begin{figure}[tbh]

\hspace{-0in}\resizebox{3.1in}{!}{\includegraphics{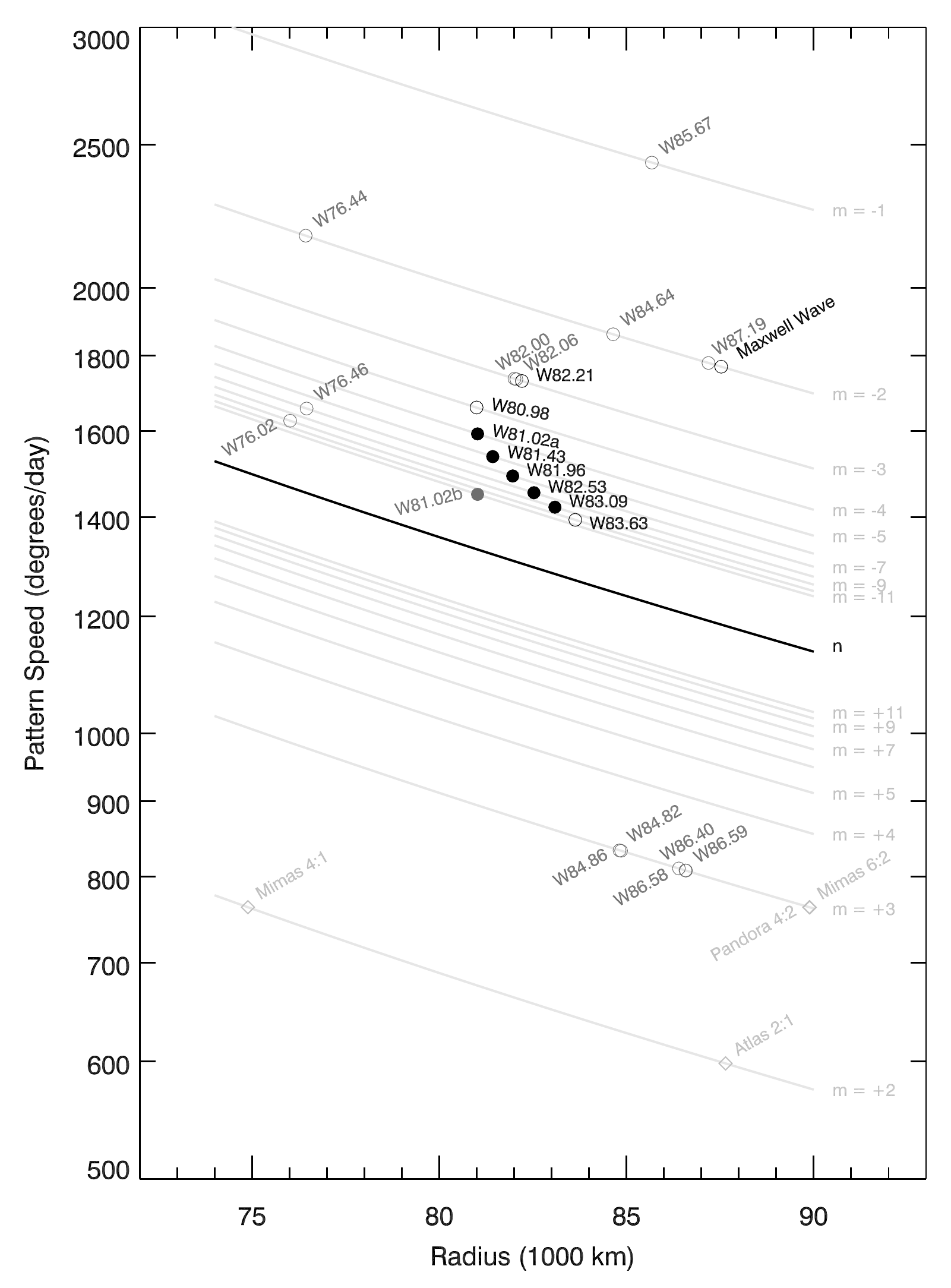}}
\caption{Summary plot showing the locations and pattern speeds of the currently-identified density waves that can be attributed to structures inside the planet. {The filled symbols are those first identified in this paper, while the open symbols are those found in previous publications \citep{HN13, HN14, French16, French18}. The black symbols correspond to the waves most likely generated by fundamental sectoral normal modes. Note that at this scale the symbols are much larger than the radial extent of the waves.}}
\label{resloc}
\end{figure}

 \begin{figure}[tbh]
\resizebox{3.1in}{4.2in}{\includegraphics{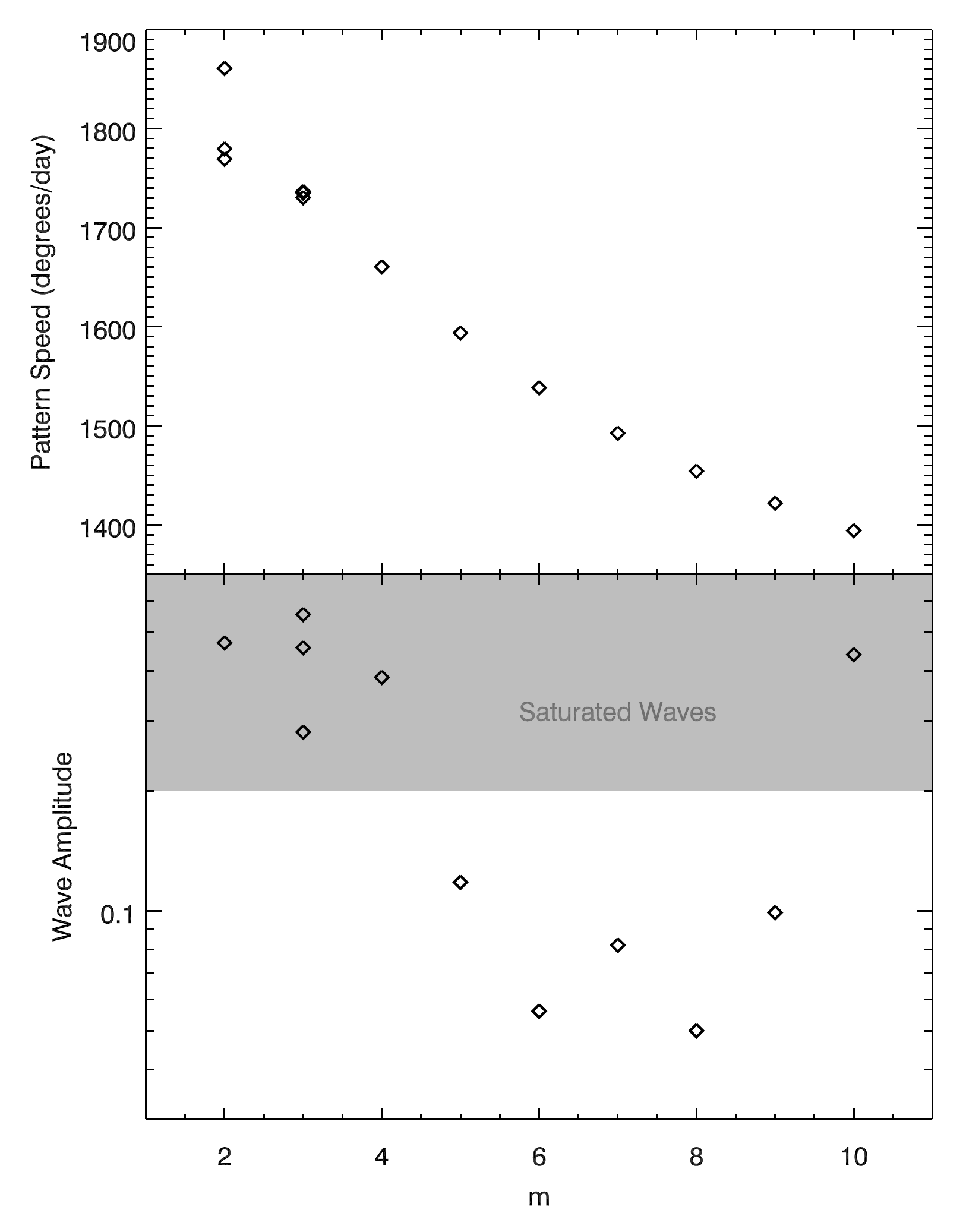}}
\caption{Summary of the waves that can be attributed to planetary  fundamental sectoral normal modes. The top panel shows the pattern speeds of the waves as a function of $m$. The bottom panel shows the wave amplitudes versus $m$. Note that waves in the grey region are probably nonlinear and somewhat saturated.}
\label{params}
\end{figure}

\section{Summary}
\label{summary}

Table~\ref{wavetab} summarizes the properties of the waves and wave candidates in the middle C ring derived from the above analyses, while Figure~\ref{resloc} shows the locations and pattern speeds of these waves, together with the other currently identified density waves in the C ring. This plot clearly demonstrates that the waves W81.02a, W81.43, W81.96, W82.53 and W83.09 are part of the same sequence as W83.63, W80.98, W82.00/W82.06/W82.21 and W84.64/W87.12/Maxwell ringlet. This is consistent with all these waves being associated with fundamental sectoral normal modes predicted by \citet{MarleyPorco93}. Having a full sequence of waves corresponding to fundamental sectoral normal modes with $m=\ell=2-10$ should be particularly informative for efforts to model Saturn's internal structure and overall rotation rate.  By contrast, W81.02b falls well off of this trend, and is therefore most likely generated by one of a different class of planetary oscillations. Most likely, this wave is generated by a planetary oscillation with $\ell=13$ and $m=11$. Such an interpretation is not only consistent with extrapolations from earlier predictions \citep{MarleyPorco93, Marley14}, but also more recent calculations of planetary normal modes \citep{M18}. {Recall that \citet{French18} identified half a dozen density and bending waves in the inner C ring, most of which are probably also generated by fundamental but non-sectoral normal modes, and so should provide further constraints on Saturn's interior \citep{M18}.} Detailed interior modeling is beyond the scope of this report, but we can note some interesting trends among the amplitudes and pattern speeds of these waves (see Figure~\ref{params}). 

First, we may note that while there are multiple waves with similar pattern speeds for $m=-2$ and $m=-3$, this is not does not appear to be the case for any of the other waves in this sequence. \citet{Fuller14} suggests that the multiple $m=-2$ and $m=-3$ waves represent mixing between the fundamental sectoral normal modes and $g$-mode waves within a stably stratified layer in Saturn's interior. While this model does predict that such ``mixed'' modes would be weaker for modes with $m=4$ than for $m=2$ or 3, our search did not uncover any weak $m=-4$ signals close to the W80.98 wave (see Figure~\ref{m2-6search}) or any weak $m=-10$ waves around W83.63  (see Figure~\ref{m7-11search}), even though W80.98 and W83.63 are not much weaker than the $m=-3$ waves. Furthermore, we did not find any additional examples of waves with $m=-2$ or $m=-3$ in this region. This may suggest that only a limited number of mixed modes are efficiently excited. If nothing else, the lack of additional waves with $m=-4$ through -10 removes many potential ambiguities in the interpretation of these oscillations.

The amplitudes of these waves also show some interesting trends. Note that these wave amplitudes are directly proportional to the mode amplitudes inside the planet, scaled by a factor of  {${2m+\ell+1}$} \citep{MarleyPorco93}. However, the conversion factor also depends upon the damping length of the wave \citep{Tiscareno07}, which is difficult to independently constrain for these weak waves. Systematic uncertainties in the damping length do not strongly affect the inferred relative amplitudes of the planetary modes, but do impact their absolute values. Hence we will only consider relative amplitudes among the observed waves here.

As shown in Figure~\ref{params}, the waves with $m=-2,-3,-4$ and $-10$ are all probably saturated, so detailed comparisons are problematic, but we can clearly see that the waves with $m=-5, -7$ and $-9$ are all significantly lower amplitude than those with $m=-2,-3,-4$ and $-10$, while those with $m=-6$ and $m=-8$ are weaker still. At first, this seems to suggest that there is a minimum in the wave amplitudes around $m=-7$. The problem with this interpretation is that we could not find any wave signature with $m=-11$ outside W83.63, where the resonance with the fundamental sectoral mode with $\ell=m=11$ should reside. This, along with the lack of additional unidentified waves outside 84,000 km, suggest that fundamental sectoral modes with $m>10$ are as weak or weaker than those with $m=5-9$. Hence it is probably more accurate to say that  all modes with $m>4$, except for $m=10$, have low amplitudes compared to those with $m=2,3$ or $4$. 

A general decrease in mode amplitude with increasing $m$ is consistent with various earlier predictions \citep{MarleyPorco93, M18}. However, these theoretical models generally predict a steady decrease in the mode amplitudes, while the data seems to show a more abrupt transition around a critical pattern speed around 1630$^\circ$/day. This pattern speed also appears to be relevant for the amplitudes and visibility of waves generated by other planetary normal modes. All of the normal-mode waves identified in the inner C ring by \citet{French18} have pattern speeds greater than 1626$^\circ$/day, while most of the remaining normal modes that have no identified waves (except for a few that may fall within gaps or close to the strong Titan apsidal resonance)  have pattern speeds below 1600$^\circ$/day \citep{M18}. Hence it appears that modes with pattern speeds greater than about 1620$^\circ$/day tend to have large enough amplitudes to produce obvious density waves, while those with pattern speeds less than this value are substantially weaker (with the exception of the $m=10$ fundamental sectoral normal mode). Interestingly, 1620$^\circ$/day is roughly twice the planet's bulk rotation rate, so perhaps this transition has some relationship to which modes can be efficiently excited inside a rotating planet. Such possibilities, along with an explanation for the exceptionally strong $m=10$ mode will need to be explored in future works.

\section*{Acknowledgements}

The authors would like to thank the VIMS team, the Cassini project and NASA for the data used in this analysis. We would also like to thank M. El Moutamid, M. Marley, C. Mankovich and E. Dederick for useful conversations. This work was supported by NASA Cassini Data Analysis Program Grant NNX17AF85G.


\end{document}